\documentclass[fleqn,usenatbib]{mnras}

\usepackage[T1]{fontenc}
\usepackage{ae,aecompl}

\usepackage{graphicx}	% Including figure files
\usepackage{amsmath}	% Advanced maths commands
\usepackage{amssymb}	% Extra maths symbols

\title[The Comptonising medium of 4U $1636-53$]{The Comptonising medium of the neutron star in 4U 1636$-$53 through its lower kilohertz  Quasi-Periodic Oscillations}

\author[K. Karpouzas et al.]{
Konstantinos Karpouzas$^{1,2}$,\thanks{E-mail: karpouzas@astro.rug.nl}
Mariano M\'endez$^{1}$,
Evandro M. Ribeiro$^{1}$,
\newauthor
\ Diego Altamirano$^{2}$, Omer Blaes$^{3}$ and Federico Garc\'ia$^{1}$
\
\\
$^{1}$Kapteyn Astronomical Institute, University of Groningen, P.O. BOX 800, 9700 AV Groningen, The Netherlands\\
$^{2}$School of Physics and Astronomy, University of Southampton, Southampton, SO17 1BJ, UK\\
$^{3}$Department of Physics, University of California, Santa Barbara, CA 93106, USA
}

\date{Accepted 2019 December 6.  Received 2019 November 11.  in original form 2019 July 20}

\pubyear{2019}

\begin{document}
\label{firstpage}
\pagerange{\pageref{firstpage}--\pageref{lastpage}}
\maketitle

\begin{abstract}
Inverse Compton scattering dominates the high energy part of the spectra of neutron star (NS) low mass X-ray binaries (LMXBs). It has been proposed that inverse Compton scattering also drives the radiative properties of kilohertz quasi periodic oscillations (kHz QPOs). In this work, we construct a model that predicts the energy dependence of the rms amplitude and time lag of the kHz QPOs. Using this model, we fit the rms amplitude and time lag energy spectra of the lower kHz QPO in the NS LMXB 4U $1636-53$ over 11 frequency intervals of the QPO and report three important findings: (i) A medium that extends 1-8 km above the NS surface is required to fit the data; this medium can be sustained by the balance between gravity and radiation pressure, without forcing any equilibrium condition. (ii) We predict a time delay between the oscillating NS temperature, due to feedback, and the oscillating electron temperature of the medium which, with the help of phase resolved spectroscopy, can be used as a probe of the geometry and the feedback mechanism. (iii) We show that the observed variability as a function of QPO frequency is mainly driven by the oscillating electron temperature of the medium. This provides strong evidence that the Comptonising medium in LMXBs significantly affects, if not completely drives, the radiative properties of the lower kHz QPOs regardless of the nature of the dynamical mechanism that produces the QPO frequencies.

\end{abstract}


%%%%%%%%%%%%%%%%%%%%%%%%%%%%%%%%%%%%%%%%%%%%%%%%%%

%%%%%%%%%%%%%%%%% BODY OF PAPER %%%%%%%%%%%%%%%%%%

\section{Introduction}
\par 
Kilohertz quasi periodic oscillations (kHz QPOs) represent the fastest variability observed in the X-ray light-curves of neutron star (NS) low mass X-ray binaries (LMXBs) to date. \citet{Doesburgh2018}, reported a QPO at 1267 Hz in the NS LMXB 4U $0614+09$, which is the highest ever observed in such a system. A QPO is  a narrow, albeit with finite width, peak in the power density spectrum (PDS) of a light-curve. See \citet{vanderKlis1989} for more information on the timing techniques. QPOs in LMXBs had been previously discovered at lower frequencies, however the timing resolution and large effective area of modern instruments made it possible to also detect them at kHz frequencies. Throughout this paper ,we refer to the kHz QPOs when the QPOs appear between approximately 400 Hz and 1200 Hz.

\par The history of kHz QPOs begins with the launch of the Rossi X-ray timing explorer \citep[RXTE]{Bradt1993}. The  first detections and analyses of the newly discovered kHz QPOs are presented in \citet{vdk1996}; \citet{vanParadijs1996}; \citet{Strohmayer1996}; \citet{Berger1996}; \citet{Zhang1996} and \citet{Ford1997}. The kHz QPOs in NS LMXBs, that were reported in the aforementioned papers, usually consist of a lower and an upper QPO, separated by a few hundreds of Hz. Later, work done by \citet{Wijnands1997}; \citet{Wijnands_2_1997}; \citet{vanderklis1997} and \citet{Mendez1998}; \citet{Mendez1999}; \citet{Zhang2006} shed more light into the existence of pairs of kHz QPOs, or twin as they are often called. For a detailed review on the first detections and characterization of kHz QPOs we refer the reader to \citet{vanderklis1998review}.

Theoretical models trying to explain the origin of the QPO frequencies date back to 1985, before kHz QPOs were even detected. Among the first attempts to link the low-frequency QPOs to a physical mechanism, was the beat frequency model, as described in \citet{Alpar1985}, where the QPO frequency is a beat between the spin of the NS and the Keplerian frequency of the accreting material. Later, \citet{Hameury1985} connected the QPO frequency with the NS spin. More specifically, a magnetic field driven mechanism was thought to be able to produce bright spots on the NS surface, which rotate at the NS spin, causing brightness oscillations to manifest themselves as QPOs. 

 \citet{Boyle1986}, assumed for the first time a spatially resolved Comptonising medium, which we will hereafter refer to as the corona, whose optical depth was linked to the QPO frequency. In the model of \citet{Boyle1986} the corona can exist on top of an accretion disc. The optical depth of this corona has a spatial dependence and thus can be linked to a Keplerian frequency associated to a certain distance inside the corona. The optical depth is regulated by the incident flux, from the central object onto to the outer surface of the corona, and for the first time is linked to the QPO frequency. The aforementioned work models the intensity of a source as a function of a Keplerian frequency, and a fit to the source GX5-1 resulted in reasonable values for the spectral parameters. Another pioneering QPO model was introduced by \citet{Fortner1989}, where the frequency of an oscillating radial flow of material is associated to radiation-hydrodynamic overstability that causes the observed QPOs.

Shortly after the discovery of kHz QPOs, the statistical properties and dependence of these high-frequency oscillations upon observed physical quantities became too complex to be explained by the models that had been proposed to explain the low-frequency QPOs. Without going into detail,  \citet{Klein1996}; \citet{Kaaret1997}; \citet{Miller1998}; \citet{Stella1998}; \citet{Titarchuk1998}; \citet{Osherovich1999} and  \citet{Zhang2004}, proposed a new generation of models that led to a more physically robust explanation of the observed kHz QPO frequencies, based on their latest observational properties (see \citet{vanderklis2006} for a review of these models).  

\par At the same time as their discovery, a lot of attention was given to the energy resolved variability of kHz QPOs. By energy resolved variability, we mean the study of how the properties of the kHz QPOs, namely the fractional root mean square amplitude (hereafter rms) and time (or phase) lags, depend on photon energy. For a review on timing techniques, and derivation of rms and lags, we refer the reader to \citet{Uttley2014}. The spectral analysis of a source, combined with the study of energy resolved variability has recently become famous as the spectral-timing approach.  

\par  \citet{Strohmayer1996} found, for the first time, that the rms of the lower kHz QPO in the LMXB 4U $1728-34$ increased with energy from 3$\%$, at energies between 2-6 keV, to $\sim$9.5$\%$ at 12 keV. Later, \citet{Berger1996} found that the rms of the $\sim 800$Hz QPO in 4U $1608-52$ increased with energy from 4$\%$ at 2 keV to $15\%$ at 12keV, while at higher energies it remained constant within the uncertainties. \citet{Zhang1996} suggested that the high values of rms, of the kHz QPOs at high energies indicate that the QPOs originate either at the NS surface or the corona, since the temperature of both is higher than that of the disc. \citet{Mendez1997M} carried out an interesting analysis of the absolute root mean square (RMS) in multiple energy bands for the source 4U $0614+09$. More specifically, the absolute RMS in units of counts/(s keV) is a measure of the shape of the oscillating spectrum that would be on top of a non-varying component (e.g. a truncated accretion disc). \citet{Mendez1997M} found that a Wien spectrum of temperature $\sim$ 1.56 keV and radius of $\sim$ 500m (at a distance of 3 kpc ) could explain the data. In the same work, the fractional rms spectrum is moderately well fitted by a black body spectrum with a $\sim$ 2.5$\%$ variation in its temperature and by a Comptonized spectrum with a $\sim$ 5$\%$ variation in its optical depth.

\par Later, \citet{Revnivtsev1999} introduced a technique called frequency resolved spectroscopy (FRS), in which the spectrum of the absolute RMS is constructed in separate frequency bands with the help of the energy resolved PDS. An application of the FRS technique was discussed by \citet{Gilfanov2003}, who showed that the spectral shape of the absolute RMS, resolved around the QPO frequency for the sources GX $340+0$ and 4U $1608-52$, is in very good agreement with the average spectrum of the source if a disc blackbody component is subtracted from the energy spectrum. These authors claimed that the remaining energy spectrum, after subtracting the disc blackbody, matches that of the boundary layer. A simple Wien spectrum, representing the boundary layer, fits their data well, but they note that, depending on the spectral state of the source, taking into account thermal Comptonisation might improve the fits significanty.    

\par At the same time, the study of the energy-dependent time lags was added to the spectral timing approach. In particular, after the discovery that, in the lower kHz QPOs, the soft photons lag behind the hard ones, an effect that we will refer to as soft lags, by \citet{Vaughan1998}; \citet{Kaaret1999}; \citet{Markwardt2000}; \citet{Avellar2013}, the study of time lags became an important part of any comprehensive spectral-timing analysis. \citet{Avellar2013}, mentioned  for the first time that the energy-dependent time lags of the upper kHz QPOs exhibit a behaviour significantly different to that of the lower kHz QPO, such that the hard photons lag behind the soft ones, producing the so called hard lags. The findings presented in \citet{Avellar2013}, suggest a different origin for the lower and the upper kHz QPO.  

Later, a detailed spectral-timing analysis of the NS LMXB 4U $1728-34$, done by \citet{Peille2015} revealed that although the rms increased with energy, both for the lower and the upper kHz QPOs, the time lags of the lower kHz QPOs were always soft whereas for the upper kHz QPO the time lags were mostly hard. These authors also fitted the RMS, or covariance, spectra of the sources 4U $1608-52$ and 4U $1728-34$ with a thermal Comptonisation model and obtained a very good fit. In agreement with \citet{Avellar2013}, \citet{Peille2015} concluded that the physical mechanisms driving the lower and upper kHz QPOs must be different and that the oscillating component in the source spectra, thus the QPO, must be associated with Comptonisation. A systematic study of multiple sources presented by \citet{Troyer2018ApJ}, also verified the differences between upper and lower kHz QPOs. However, the question of whether the kHz QPOs originate in the accretion disc, the NS surface, the NS boundary layer, or in a surrounding Comptonising region remained unknown. 

The explanation of the dependence of the time lags and rms upon photon energy is still highly debatable. Predicting the energy dependence of timing properties such as time lags, due to Comptonization, dates back to \citet{Wijers1987}, who used a Monte Carlo simulation to calculate the Comptonization of a black-body spectrum inside a spherical homogeneous cloud, highlighting the importance of energy dependent time-delay measurements as a probe for both the source of soft photons and the properties of the Comptonizing cloud. Specific models trying to explain this dependence were introduced by \citet{Lee1998}, \citet{Lee2001} and \citet{Kumar2014}. The models described in the aforementioned papers assume that the flux oscillation of the time averaged spectrum, that is modulated at the QPO frequency, is produced by an oscillation in the thermodynamic properties of a Comptonising region, hereafter the corona, or of the seed photon source, which can be either the NS surface, accretion disc, or both. More specifically, these thermodynamic properties are the temperature and external heating rate of the corona, the electron density of the corona and finally the temperature of the soft photon source that provides the seed photons for Comptonisation. 

\citet{Lee1998} assumed that, if thermal Comptonization is responsible for the shape of the rms spectrum (\citealt{Mendez1997M,Gilfanov2003,Peille2015}), then thermodynamic properties such as the heating rate and temperature of the corona and seed photon source should be also oscillating. The simplest case would be one in which the latter quantities oscillate exactly at the QPO frequency. The novel idea presented in \citet{Lee1998} predicts the rms and time lags due to Comptonisation, of a seed photon source in a homogenous and spherically symmetric corona. Temperature oscillations of either the corona or the seed photon source, but also the effect of an oscillating electron density, are used to reproduce the rms and time lag spectra. The authors apply an empirical fit to rms and time lag spectra of the source 4U $1608-52$, with a model that uses a superposition of the three oscillations mentioned above, and conclude that a time varying coronal electron density is more likely to drive the observed behaviour. They also find a size of around 6 km for the, assumed spherical, corona, which reinforced the idea that the kHz QPOs are either produced or radiatively enhanced in a region close to the NS.

The hard time lags of the source 4U $1608-52$ initially reported in \citet{Vaughan1997ApJ}, that were used by \citet{Lee1998}, were later on proven to have the opposite sign. After \citet{Vaughan1998} clarified that the time lags of 4U $1608-52$ were actualy soft, a second work by \citet{Lee2001} assumed that an oscillation in the temperature of the corona, that produces the QPO, causes a delayed oscillation in the temperature of the seed photon source through feedback. By feedback we mean that a fraction of the photons Comptonised in the corona return to the seed photon source through random walks, and re-heat it. 
Because of this delayed response of the seed photon source, the low energy photons are expected to lag behind the hard ones, thus explaining the observed soft lags. However, in \citet{Lee2001} the QPO amplitude is approximated as a linear combination of the amplitudes of the corona and seed photon source oscillations. This ad-hoc method is used in order to connect the two amplitudes (corona and seed photon source) through feedback and calculate one as a function of the other. This approach succeeds in explaining, at least qualitatively, the behaviour of the rms and time lags. However, the major disadvantage of this model is that the energy balance of the corona, through Compton cooling, is not taken into account and thus the amplitude of the temperature oscillations do not emerge as natural consequences of a self consistent physical process.

A self consistent model, that takes into account the dynamic cooling of the corona and the feedback onto the soft photon source was proposed by \citet{Kumar2014}. Since the work presented here uses, and builds upon, the aforementioned model we will provide a detailed description of the model in the next section. \citet{Kumar2016} used this model for a preliminary study of the rms and lag energy spectra of the NS LMXB 4U $1608-52$. The authors found that a small corona ($\sim$2-6 km) and a significant amount of feedback (20-80$\%$) are required to qualitatively explain the observed behaviour. Finally, \citet{Kumar2016MC} used a Monte Carlo method to estimate the expected amount of photons that would impinge back onto the seed photon source during Comptonisation. They simulated different corona geometries and concluded that in all cases a significant amount of feedback is justified.

\par Although self consistent, the model of \citet{Kumar2014} failed to predict the flattening of the rms energy spectrum above 12 keV that was observed by \citet{Berger1996} in the NS LMXB 4U $1608-52$. Attempts to explain the aforementioned flattening were made before, initially by \citet{Miller1998} and later by \citet{Lee1998}, based on the assumption that the optical depth is oscillating at the QPO frequency. The assumption of an oscillating electron number density, and thus optical depth, allowed the model of \citet{Lee1998} to explain the flattening of the rms energy spectrum better. However, to explain the soft time lags of 4U $1608-52$ the model of \citet{Lee2001} introduced the assumption of feedback and an oscillation in the temperature of the corona instead of an oscillating optical depth. Particularly interesting is the NS LMXB 4U $1636-53$, for which spectral analyses date back to 1986 (see \citealt{Vacca1986}; \citealt{Tatsumi1987} and \citealt{Damen1990}). Recently, \citet{Ribeiro2019} showed that the rms energy spectrum of the NS LMXB 4U $1636-53$ depends highly on the lower kHz QPO frequency and, on average, decreases at energies above 12 keV, a fact that might present an interesting challenge for all of the aforementioned models.

\par Evidence that the properties of the corona undergo changes was already given by \citet{Barret2013}. In the aforementioned work the optical depth, corona temperature and seed photon temperature, of the NS LMXB 4U $1608-52$, appear to have a dependence on QPO frequency. A similar dependence was reported, for the NS LMXB 4U $1636-53$, in \citet{Zhang2017} and \citet{Ribeiro2017}. Furthermore, phase resolved spectroscopy of type B QPOs in the black hole X-ray binary GX $339-4$ and of kHz QPOs in the NS LMXB 4U $1608-52$, presented in  \citet{Stevens2016} and  \citet{Stevens2018cosp...42E3258S}, respectively, revealed that the corona temperature and seed photon source temperature oscillate at the QPO frequency. Preliminary results (\citealt{Stevens2019AAS}) indicate that the properties of the corona and soft photon source, apart from oscillating coherently at the QPO frequency, show systematic time lags between them. All of the the above results prove the importance of linking the variability of the photon number density (QPO) to the spectral and geometric properties of the corona. 

In the present work we reproduced the model of \citet{Kumar2014} with minor changes in the physical assumptions and a basic change of the solving scheme. These changes are discussed in section 2, with the help of Appendix A. In section 3, we discuss the process of fitting our model to a set of frequency resolved rms and time lag spectra of the lower kHz QPO, in the NS LMXB 4U $1636-53$ (\citealt{Ribeiro2019}). Finally, in sections 4 and 5 we summarize our results and discuss our findings on the magnitude of the measured quantities and their dependence upon frequency.

\section{The model}
\label{sec:model}

In this section we give a brief description of the model presented in \citet{Kumar2014} alongside the changes introduced by us. The model assumes that the surface of a neutron star of radius $a$ and temperature $T_s$, injects photons inside a spherically symmetric medium of width $L$, consisting of highly energetic electrons, the so called corona, centered at the NS, with a rate $\dot{n}_{s\gamma}$. The photons are inverse Compton scattered and, since the medium is finite, after a photon of energy $E$ undergoes certain number of scatterings it will escape the medium with an energy $E'>E$. The escape rate per unit volume is $\dot{n}_{esc}$. This symmetry, and the assumption that the density of the corona is uniform, simplifies the definition of the (Thomson) optical depth, which reduces to $\tau_T=\sigma_{T}n_eL$, where $\sigma_T$  is the Thomson cross section and $n_e$ the electron number density of the plasma. \citet{Schulz1993} showed that theoretical spectra constructed with a simple geometry, such as the one described above, can yield good fits to the data.

\par The time dependent evolution of the spectrum, $n_{\gamma}(E,t)$, that undergoes multiple inverse Compton scatterings is described by the Kompaneets equation \citep{kompaneets1957}. We use the form of the equation that was first presented in \citet{Psaltis1997}, then used by \citet{Lee1998}, \citet{Lee2001} and later by \citet{Kumar2014}. This equation can be written as:

\begin{equation}
\begin{split}
t_c \frac{\partial{n_{\gamma}}} { \partial{t} } = \frac{1}{m_e c^2} \frac{ \partial{} }{ \partial{E} } \Big( -4kT_eEn_{\gamma} + E^2 n_{\gamma} +kT_e\frac{ \partial{} }{ \partial{E} }(E^2n_{\gamma}) \Big) \\ + t_c\dot{n}_{s
\gamma} - t_c
\dot{n_{esc}} \ \ ,
\label{eq:Kompaneets}
\end{split}
\end{equation}

where  $t_c=L/c\tau_T$ is the Thomson collision time scale, $T_e$ the temperature of the medium, $k$ the Boltzmann constant, $m_e$ the electron rest mass and $c$ the speed of light. Assuming that the soft photon source emits like a blackbody, the photon injection rate, $\dot{n}_{s\gamma}$, is  

\begin{equation}
\dot{n_{s \gamma}} = \Big[ \frac{3 a^2}{ [ (a+L)^3 - a^3 ] } \Big] \Big( \frac{2 \pi}{h^3c^2}  \frac{E^2}{e^{\frac{E}{kT_s}} -1} \Big) \ \ , 
\label{eq2:seed}
\end{equation}

\noindent where $h$ is the Planck constant and $a$ the NS radius. We used the same notation and symbols as in \citet{Kumar2014} so far in order to avoid confusion and built upon their model in a coherent way. Equation (\ref{eq:Kompaneets}) is subject to specific assumptions, some of which will be summarised now for the sake of clarity. First of all, a uniform and isotropic distribution of electrons and photons is assumed. We thus ignore the spatial dependence of the photon phase-space density and also ignore any effects of electron bulk motion inside the plasma. The seed photon source is the NS surface which, combined with the fact that the medium is a spherically symmetric shell, means that each photon is emitted at equal distance from the surface of a sphere with radius $L+a$. 

\par  We further assume that photons escape the medium at a constant rate, $\dot{n}_{esc}$. According to equation (\ref{eq:Kompaneets}), the number density of photons with energy between $E$ and $E+dE$ at a given time is $n_{\gamma}(t,E)$. After production, a photon undergoes a number of scatterings, $N_{esc}$, before it escapes from the medium. Assuming that each scattering is an independent event, we can assign an escape probability per scattering, $P_{esc}=1/N_{esc}$. For $N_{esc}$ we used the same formulation as in \citet{Lightman1987}, which is based on the solution of the photon diffusion equation, under the assumption that the intensity of the source inside the sphere is sinusoidally distributed with radius, as presented in \citet{Sunyaev1980}. Using the Klein-Nishina optical depth, $\tau_{KN}$, we can write, 

\begin{equation}
\begin{split}
 N_{esc}= \tau_T + \frac{1}{3} \tau_T \tau_{KN}(E)f(E),
\label{eq:Nesc}
\end{split}
\end{equation}

\noindent where

\begin{equation}
   f(E)=  
   \begin{cases}
      1 &{\rm if} \quad E \leq 0.1 m_e c^2 \cr
      (1-\frac{E}{m_e c^2})/0.9 &{\rm if} \quad 0.1 m_e c^2 <E<m_e c^2 \cr 
       0 &{\rm if} \quad E \geq m_e c^2  
   \end{cases}
\label{fcorr}
\end{equation}

\noindent and given the Klein-Nishina cross section, $\sigma_{KN}(E)$, the Klein-Nishina optical depth is 

\begin{equation}
\tau_{KN}(E) = \frac{\tau_T \sigma_{KN}(E)}{\sigma_T}.
\label{eq:klein}
\end{equation}

In order to calculate $\dot{n}_{esc}$, which according to equation (\ref{eq:Kompaneets})  has units of photons per unit volume per unit energy per unit time, we must multiply $n_{\gamma}(E,t)$ by the escape probability density function, which is assumed as a constant number that represents probability per unit time. Thus, based on the fact that the average time a photon takes to escape the medium is $t_c N_{esc}$, the escape probability per unit time is $\frac{1}{t_{esc}N_{esc}}$ or $\frac{c \tau_T}{L N_{esc}}$  so that

\begin{equation}
\dot{n}_{esc}=\frac{c \tau_T n_{\gamma}(t,E)}{L N_{esc} } = \frac{c}{L} \frac{n_{\gamma}(t,E)}{ \Big[ 1 + \frac{1}{3}  \tau_{KN}(E)f(E)  \Big]}
\label{eq:nesc}
\end{equation}

Equation (\ref{eq:nesc}) is the same as equation (21a) of \citet{Lightman1987}. We must note that in \citet{Lightman1987} the authors include a relativistic correction term, $\omega(x)$, in equation (22a) of their paper. Given that the majority of the spectral analyses presented in the literature use $nthcomp$ (\citealt{Lightman1987,Zdziarski1996}) to model thermal Comptonization, we want our model for $n_{\gamma}$ to resemble that of $nthcomp$ as much as possible for a given set of spectral parameters. Therefore, in our approach for $\dot{n}_{esc}$ we include Klein-Nishina corrections and the same form of $N_{esc}$ that is used in \citet{Lightman1987}, because the aforementioned paper is the basis of the $nthcomp$ model (\citealt{Zdziarski1996}) in $XSPEC$ (\citealt{§Arnaud1996ASPC}). To conclude with the assumptions, the electrons of the plasma are hot, but not relativistic ($kT_e<<m_ec$) in our case, thus allowing for the use of the Thomson cross section, $\sigma_T$, for the scatterings. In spite of this, we use the Klein-Nishina cross section, $\sigma_{KN}$, to preserve generality throughout this work, but we note that using $\sigma_T$ would not change any of the results of this work, because we mainly work with energies $E<<m_e c^2$. The steady state solution ($\frac{\partial{n_{\gamma}}}{\partial{t}}=0$, hereafter SSS) of equation (\ref{eq:Kompaneets}) is the energy averaged spectrum, $n_{\gamma 0}$, i.e. the Comptonised continuum of the observed spectrum. Equation (\ref{eq:Kompaneets}) is discretised and solved numerically as a boundary value problem in order to obtain $n_{\gamma 0}$. In Appendix \ref{ap:math} we provide the reader with details on the solving scheme.

\par The model that we reconstruct in this work is based on the idea that a QPO is modeled as a small oscillation of the SSS. Regardless of whether the QPO is produced on the NS or accretion disc, or whether it is the result of oscillations of thermodynamical and physical properties of the corona, a perturbation of the SSS can be linked to a perturbation in any physical quantity that appears in equation (\ref{eq:Kompaneets}). This concept was first introduced by \citet{Lee1998}. We can write the perturbation as $n_{\gamma} = n_{\gamma0}(1 + \delta n_{\gamma} e^{-i\omega t})$, where $\omega$ is the mean frequency of the QPO and $\delta n_{\gamma}$ the complex fractional amplitude of the QPO. We further assume that this perturbation can be linked to a perturbation in the temperature of the plasma, $T_e = T_{e0}(1+\delta T_e e^{-i\omega t})$, that could be a product of an oscillating external heating rate, $\dot{H}_{ext}$, which we write as $\dot{H}_{ext} = \dot{H}_{ext0}(1+\delta \dot{H}_{ext} e^{-i\omega t})$. The external heating rate is required in order to account for the observational fact that the cooling of the medium, through repeated scatterings, does not lead to an extreme drop in the temperature of the medium. Finally it is assumed that a fraction, $\eta$, of the Comptonised photons impinge back onto the surface of the NS causing an oscillation in its temperature, $T_s = T_{s0}(1+\delta T_s e^{-i \omega t})$. In principle the NS, as a black body, could undergo inherent oscillations on its own, on top of the oscillations caused by this feedback loop.

\par All of the perturbations are written in the form $X_i = X_0(1+\delta X_i e^{-i \omega t})$ so that $\delta X_i$ is a fractional variation of the equilibrium value, $X_0$, and thus dimensionless. We must also note that in the above, $\delta \dot{H}_{ext}$, $\delta T_e$, $\delta T_s$ and $\delta n_{\gamma}$ are complex quantities, while $\delta n_{\gamma}$ is a function of energy. After linearising equation (\ref{eq:Kompaneets}) we can form a new differential equation for the complex amplitude. For the new variables $N = n_{\gamma}/n_c$ and $x = E/kT_e$ (see Appendix \ref{ap:math}), the linearised form is: 

\begin{equation}
\begin{split}
\frac{d^2 \delta n_{\gamma}}{dx^2} = -\frac{d \delta n_{\gamma}}{dx} - \frac{2}{N}\frac{dN}{dx}\frac{d\delta n_{\gamma}}{dx} + \frac{1}{N}\frac{\delta n_{\gamma}}{e^{xT_e/T_s}-1} \\ - \frac{i c_5 \delta n_{\gamma}}{x^2} + \delta T_e \Big(\frac{2}{x^2} -\frac{1}{N} \frac{d^2N}{dx^2} \Big) - \\ - \delta T_s \frac{1}{N} \frac{T_{e0}}{T_{s0}} \frac{x}{\big( e^{xT_{e0}/T_{s0}}+e^{-xT_{e0}/T_{s0}}-2 \big)} \ \ ,
\label{eq:CODE}
\end{split}
\end{equation}

\noindent where $c_5 = \frac{\omega m_e t_c c^2}{kT_{e0}}$. The complex amplitudes $\delta T_e$, $\delta n_{\gamma}$ and $\delta \dot{H}_{ext}$ are connected through the first law of thermodynamics, and $\delta T_s$ is related to $\delta n_{\gamma}$ through a feedback term added to the blackbody spectrum such that the energy balance of both the corona and seed photon source are taken into account (for more information see Appendix \ref{ap:pert}).

\par In Appendix \ref{ap:pert}, we show that $\delta T_s$, $\delta T_e$ are integral functions of $\delta n_{\gamma}$ but are also present in equation (\ref{eq:CODE}), which we need to solve in order to find $\delta n_{\gamma}$. The solution of equation (\ref{eq:CODE}), as presented in \citet{Kumar2014}, requires an initial assumption both for  $\delta T_e$  and also $\delta T_s$. The authors use the assumed values to calculate an initial solution for $\delta n_{\gamma}$, which they afterwards use to update the values of $\delta T_e$  and $\delta T_s$. This process is repeated a few times until the temperature amplitudes converge. The final solution, $\delta n_{\gamma}$, after the required steps for convergence, is taken as the final QPO complex amplitude whose modulus, $|\delta n_{\gamma}(E)|$, and argument, $\tan^{-1} {\frac{Im(\delta n_{\gamma})}{Re(\delta n_{\gamma})}}$, will translate directly into rms and phase lag respectively. If the external heating rate is assumed variable, then $\delta \dot{H}_{ext}$ can be factored out of equation (\ref{eq:CODE}) and its modulus can be used as a normalization for $| \delta n_{\gamma}$|, which is the rms. Thus, the argument, or phase, of $\delta \dot{H}_{ext}$ could be arbitrary. In Appendix \ref{ap:pert} we discuss a different mathematical approach to solve for $\delta n_{\gamma}$, $\delta T_e$ and $\delta T_s$ without the need of a converging iterative scheme. Finally, we must note that although our numerical solution, $\delta n_{\gamma}$, is calculated at each energy value of the selected grid, the actual quantity that we compare to the data, eventually, is the ratio of the integrals  $\int \delta n_{\gamma} n_{\gamma0} dE / \int n_{\gamma0}dE$ where the integration is performed within the selected energy band of the corresponding measurement.

\section{Application to the lower kHz QPO of the NS LMXB 4U $1636-53$}

\subsection{Data overview}
\label{sec:data_overview}

Recently, \citet{Ribeiro2019} presented a detailed analysis of the timing properties of the lower and upper kHz QPOs in the NS LMXB 4U $1636-53$. In particular, they studied the rms of both the lower and upper kHz QPO as a function of photon energy and frequency. In order to get a good signal to noise ratio for the rms measurements, \citet{Ribeiro2019} averaged RXTE observations with lower kHz QPOs, over narrow frequency intervals centered at frequencies that span from 530 Hz to 940 Hz. The observations that they averaged were taken at different times, but correspond always to the transition from hard to soft state (\citealt{Zhang2017}). Furthermore, \citet{Avellar2013} studied the time lags as a function of energy, in 4U $1636-53$, at different QPO frequencies and they concluded that the time lags of the lower kHz QPO were soft and did not depend on QPO frequency. 

\citet{Ribeiro2019} found that, in all frequency intervals, the rms of the lower kHz QPO of 4U $1636-53$ was an increasing function of energy up to 12 keV and then, in some cases, it decreased. The authors therefore refer to the critical energy of 12 keV as the break energy, $E_{break}$. In the same work it was also noted that the slope of the increasing ($E<E_{break}$) part, of the rms as a function of energy, initially increased with increasing QPO frequency, until a critical QPO frequency of $\sim$770 Hz while for higher frequencies the slope decreased. The frequency at which the slope is maximum coincides, more or less, with the frequency at which the energy averaged rms is also maximum. \citet{Ribeiro2019} also report that the energy averaged rms, measured for the full energy band (nominally 2-60 keV), exhibits the same dependence with frequency as the slope of the rms versus energy for the lower kHz QPO. To conclude with the data overview, for every frequency interval, defined in \citet{Ribeiro2019}, we fit the energy dependent rms, given in the aforementioned work, and the energy dependent time lags, given in \citet{Avellar2013}, which according to their results are same for each one of the selected frequency intervals.

\subsection{Fitting the model}
\label{sec:fitting}

The model has eight free parameters, namely the size, $L$, of the corona, the amplitude of the external heating rate, $\delta \dot{H}_{ext}$, the feedback parameter, $\eta$, the corona temperature, $T_e$, the Thomson optical depth, $\tau_T$, the seed photon source (NS here) temperature, $T_s$, the NS radius, $a$, and the frequency of the QPO, $\nu_{low}$. The spectral parameters $T_e$,$\tau_T$ and $T_s$ are the steady state values corresponding to the time averaged spectrum. In principle any fit of our model to rms and time lag spectra should fit the time averaged spectrum at the same time, so one would get a good constraint on the spectral parameters. In this work, since the rms and time lag spectra were derived by combining observations and since the time averaged model assumed in \citet{Zhang2017} is different than ours, we do not fix $T_e$,$\tau_T$ and $T_s$ to the values reported in \citet{Zhang2017} and \citet{Ribeiro2017}, but instead treat them as free parameters and expect to retrieve reasonable values for them. We kept the NS radius, $a$, fixed at 10 km in every frequency interval. We note that any value of the NS radius, in the 8-18 km range, does not change the results presented here except for the external heating rate, $\dot{H}_{ext0}$, which increases almost by a factor of 8 at $a=18$ km.

Equation (\ref{eq:disc}) is used to describe the total, blackbody, luminosity of the seed photon source as a superposition of a black body with an inherent temperature $T_{s,inh}$ and a feedback dominated term that is associated with an assumed fraction of photons that would fail to escape the medium and impinge back onto the NS increasing its temperature. For a set of spectral parameters $T_s$, $T_e$, $\tau_T$ and size $L$, if feedback dominates the luminosity of the soft component, we can assume the term associated with $T_{s,inh}$ on the right hand side of equation (\ref{eq:disc}) is zero and solve for $\eta$ to find the maximum feedback, $\eta_{max}$, that is allowed. Clearly, the value of $\eta_{max}$ depends on the parameters ($T_s$, $T_e$, $\tau_T$ and size $L$) and, therefore, if we use $\eta$ as a free parameter we would have to dynamically change its upper value (through a uniform prior distribution) based on the values of the other free parameters. Instead of using $\eta$ directly, we define a feedback fraction, $f_{\eta}=\eta/\eta_{max}$, which now takes values between zero and unity, and use this as a free parameter, so that the model calculates $\eta$ internally using $f_{\eta}$ and the rest of the free parameters. Since we fit our model to the data of the energy dependent rms and time lag simultaneously, we define two different likelihood functions, one for the rms and one for the time lag. Assuming that the rms and time lag are independent measurements, we define our best fit as the set of parameters that minimize the product of the two likelihoods. 

We use the python implementation of the affine-invariant ensemble sampler, introduced by \citet{Goodman2010CAMCS}, as presented in \citet{emcee2013PASP}, to find the best-fitting parameters given the data. We used 200 walkers in the 6-dimensional parameter space of $T_e$, $T_s$, $\tau_T$, $\delta \dot{H}_{ext}$, $\eta$ and $L$, and allowed 1000 iterations, or walks, for each walker on average. The QPO frequency, $\nu_{low}$, in our model was fixed to the mean of every frequency interval of the data. After performing a 20-30$\%$ burn-in on the total number of iterations, we noticed that in all frequency intervals the posterior distributions were bi-modal. In order to distinguish between the two modes we chose a pair of free parameters, that were different at each frequency interval, for which the two modes in their posterior were well separated. After a pair was chosen, we applied the Density Based Spatial Clustering Algorithm of applications with noise, DBSCAN, \citep{Ester1996ADA}. After the clusters were detected in the two dimensional density plot of a pair of parameters we assigned an ID to each member of each cluster. The IDs that we assigned correspond to the position in the parameter space of a certain walker at a specific iteration. By identifying the cluster, in any pair of parameters, and assigning IDs to their members, one can distinguish between the modes in the posterior of any parameter simply by isolating the IDs of a single cluster. We chose DBSCAN as a clustering algorithm for two reasons: The first, and most important, reason is DBSCANs ability to deal with noise-like patterns, that contaminate the two dimensional density plot of a pair of two parameters, caused by the fact that at the end of the MCMC iterations not all of the walkers have converged necessarily. The second reason is that DBSCAN also works well in cases where the clusters are not of the same size, which also happens in our case. We took as  the best-fitting value of each parameter the median of the corresponding posterior of each mode. The median was used because the posterior distributions were not strictly normal. Finally, we note that in the cases where the data value of the rms, in a certain energy band, was taken to be zero (due to a large uncertainty as specified in \citet{Ribeiro2019}, we excluded the particular data point from the fit. To get an estimate of the quality of our fits, we computed a common reduced $\chi^2$ by combining the rms and time lag measurements at each frequency and for each one of the two modes, denoted by $\chi_{m}^2$. We used the general formula $\chi_{m}^2=\frac{\sum_{i=1}^D (d_i-e_i)^2}{D-6}$, where $D-6$ is the total number of measurements of rms and time lag that were used in the fitting process minus the six degrees of freedom of our model, $d_i$ and $e_i$ are, respectively, the data and the model estimation. The subscript $m$ denotes which mode of the posterior was used.

To summarise, at each frequency interval we have two best-fitting models, one for each one of the two modes of the posteriors. This bimodality arises from the fact that, although we use the time averaged spectrum to generate the rms and time lag spectra, we do not directly fit it. As we discuss in the next section, one of the two modes predicts a low temperature for the seed photon source, $T_s$, while the other one predicts a significantly higher temperature. Therefore, following  \citet{Kumar2016} and \citet{Kumar2016MC}, we will refer to one mode as the cold-seed model and to the other one as the hot-seed model.

\section{Results }
\label{sec:results}

\begin{figure*}
	\includegraphics[width=16cm,height=21cm]{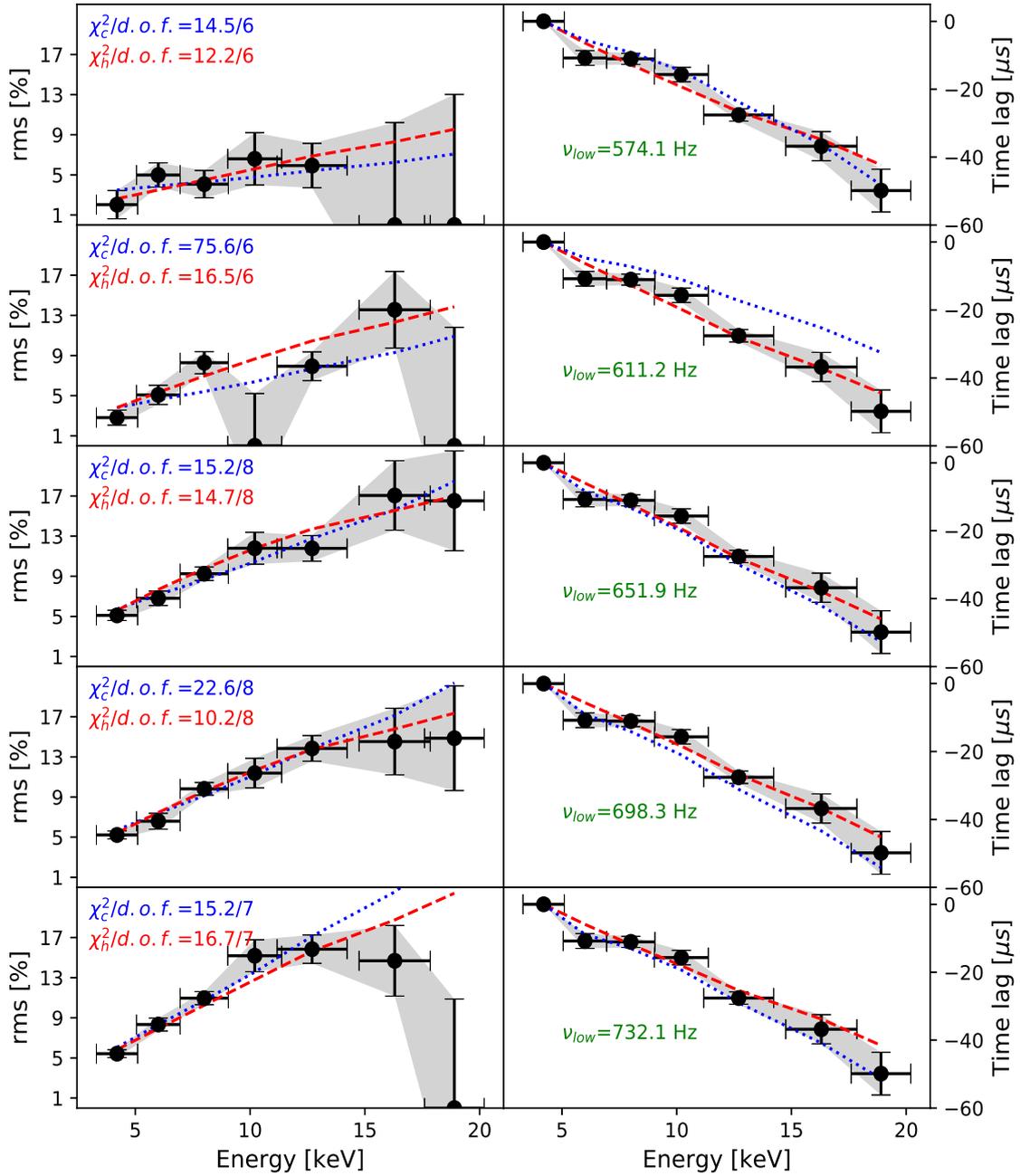}
    \caption{Best fit of the cold- and hot-seed model to the lower kHz QPO rms and time lag spectra of 4U $1636-53$, 
    in the frequency range 574.1-732.1 Hz. In the left hand side panels we plot the measured rms as 
    a function of energy (grey circles and shaded area; taken from \citet{Ribeiro2019}
     alongside our best model fits. The red dashed lines represent the best-fitting hot-seed model, 
     whereas the blue dotted lines represent the best-fitting cold-seed model. In the right hand side
      panels we plot the time lag in $\mu s$
       (grey circles and shaded area; following
         \citealt{Avellar2013}, 
         we assumed that the time lags are independent of QPO frequency) 
         alongside the
          best-fitting 
         models, 
         in the same way as
          in the left hand side. 
          }
    \label{fig:fit_flow}
\end{figure*}

\begin{figure*}
	\includegraphics[width=16cm,height=21cm]{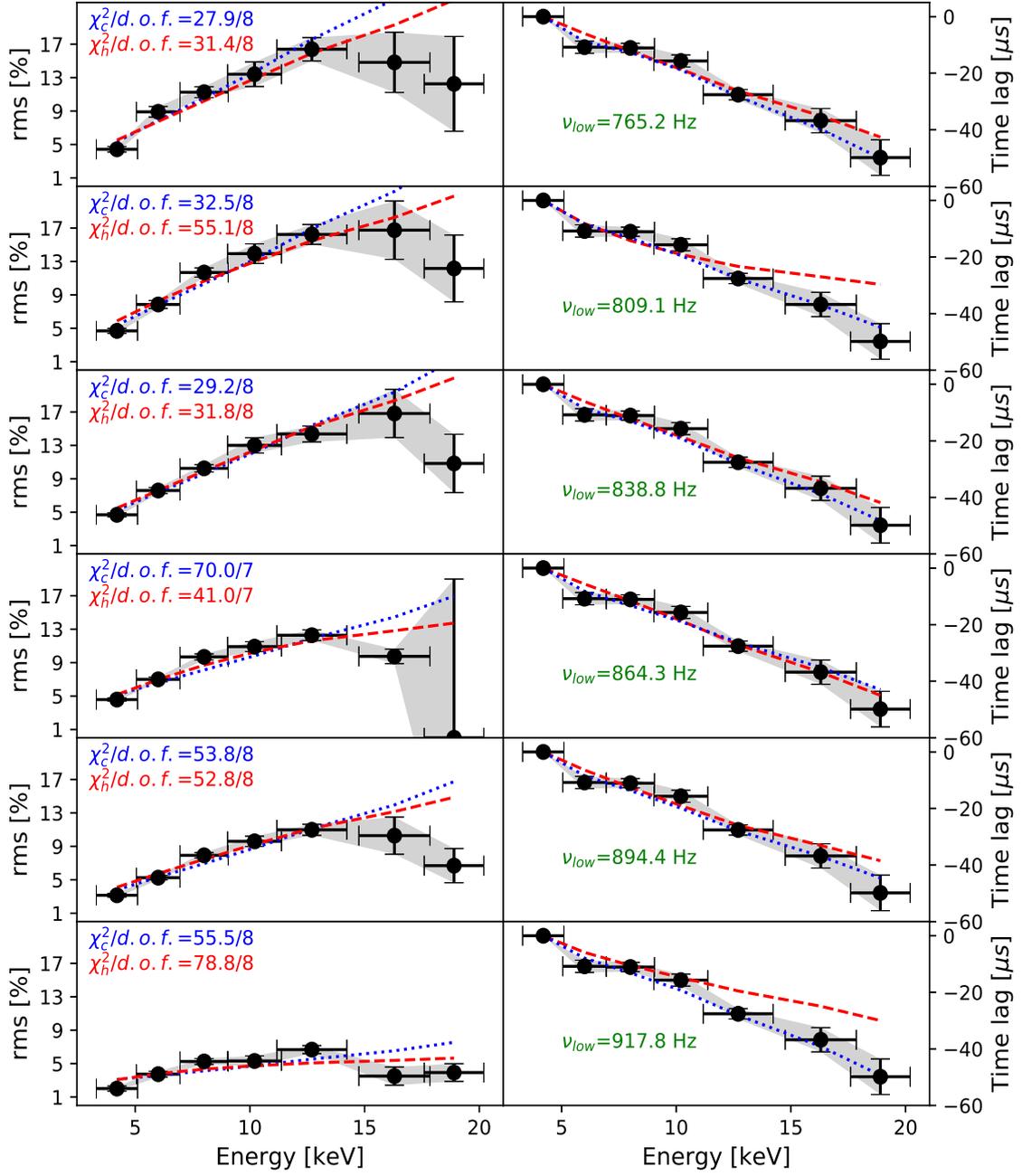}
    \caption{ Same as Figure \ref{fig:fit_flow} for the lower kHz QPO in the frequency range 765.2-917.8 Hz}
    \label{fig:fit_fhigh}
\end{figure*}

\begin{figure*}
	\includegraphics[width=17cm,height=21cm]{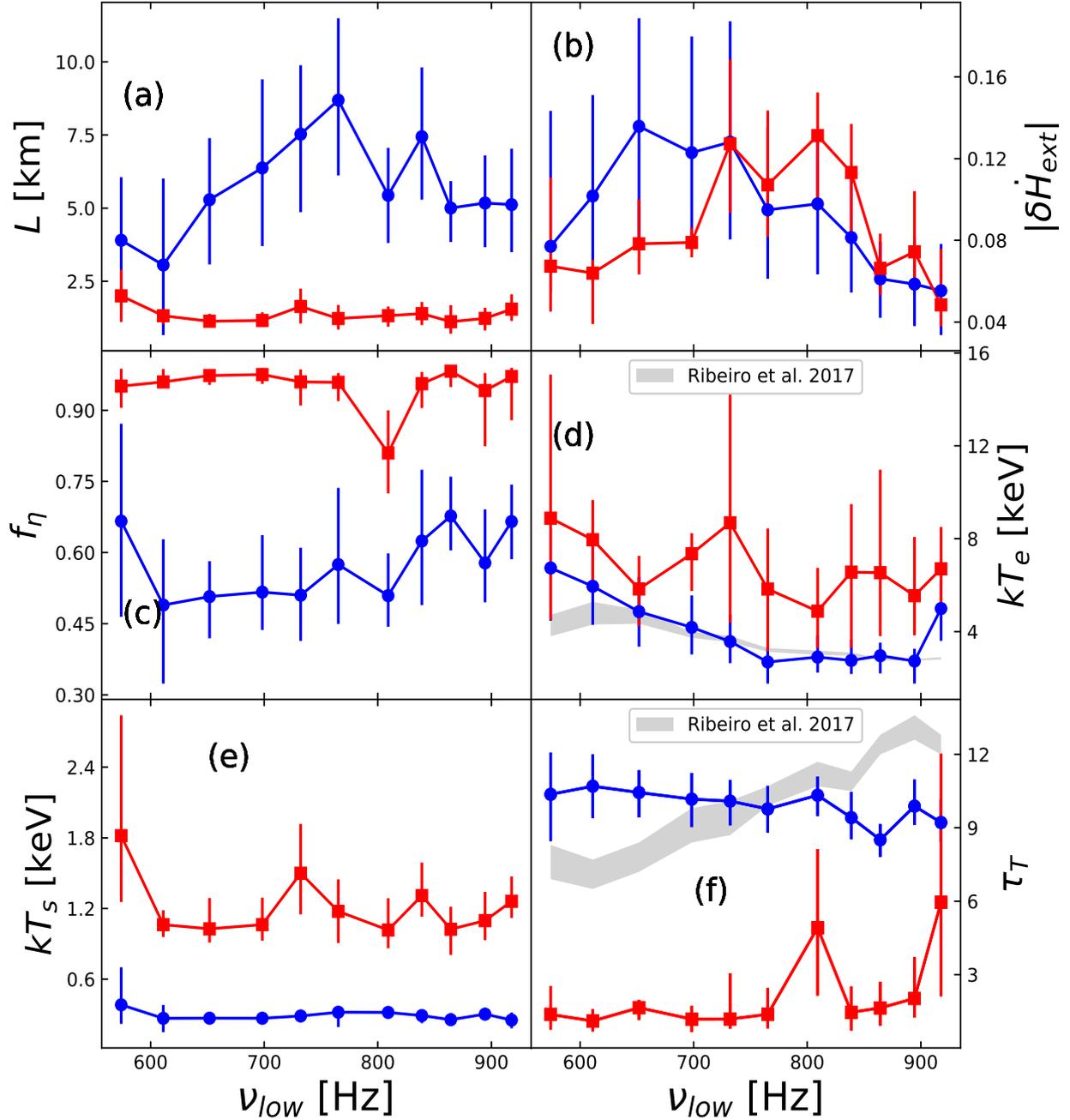}
    \caption{Summary of the MCMC results for the fits to the rms and time lag energy spectra of the lower kHz QPO in 4U $1636-53$ at all QPO frequency intervals. We plot the best-fitting parameters of the size, $L$, amplitude of the external heating rate, $|\delta\dot{H_{ext}}|$, feedback fraction, $f_{\eta}$, corona temperature, $kT_e$, NS temperature, $kT_s$, and optical depth, $\tau_T$, in panels (a), (b), (c), (d), (e) and (f), respectively, using blue squares for the cold-seed mode and red circles for the hot-seed model, alongside their 1-$\sigma$ MCMC uncertainties, as a function of frequency of the lower kHz QPO. In the middle and bottom panels, alongside our best-fitting values of $kT_e$ and $\tau_T$, we plot the same best-fitting parameters (grey shaded area) obtained by \citet{Zhang2017}, through a slightly different spectral model, as presented in \citet{Ribeiro2017}. }
    \label{fig:MCMC_cornerplot}
\end{figure*}

In Figures \ref{fig:fit_flow} and \ref{fig:fit_fhigh}, we plot our best-fitting cold- and hot-seed models together with the data of 4U $1636-53$ (\citealt{Ribeiro2019}) at each frequency interval. As we described in the previous section, we fit the model assuming that the external heating rate, $\dot{H}_{ext}$, is perturbed with a fractional amplitude $|\delta \dot{H}_{ext}|$. In table \ref{tab:fit}, we summarise the results of our model fit to every frequency interval and in Figure \ref{fig:MCMC_cornerplot}, we plot the best-fitting parameters versus the frequency of the lower kHz QPO. In each panel of the aforementioned figure we plot the best-fitting value of each one of the free parameters of both our cold- and hot-seed model, alongside the corresponding $1\sigma$ MCMC uncertainty, at each frequency interval. The asymmetric uncertainties reflect the non-symmetric, in some cases, posterior distributions.

The size, $L$, of the corona is larger in general for the cold-seed model, with a minimum of about 3.9 km at 574 Hz and a maximum of 8.7 km at 765 Hz. In the hot-seed case, the size is systematically smaller, staying below 2 km at almost all frequency intervals. The amplitude of the external heating rate, $\delta \dot{H}_{ext}$, for the cold-seed model increases with QPO frequency from approximately 8 $\%$ at 570 Hz, to a maximum of over 10$\%$ at 730 Hz and then decreases again to around $5\%$ at 890 Hz. In the hot-seed case although $\delta \dot{H}_{ext}$ has the same behaviour as in the hot-seed case, the  maximum value of 13$\%$ is found at around 800 Hz. Nevertheless, the relatively large uncertainties of $\delta \dot{H}_{ext}$, in the cold-seed case compared to those of the hot-seed case, do not allow us to place a tight constraint on the exact lower kHz QPO frequency at which $\delta \dot{H}_{ext}$ exhibits its maximum value. The feedback fraction, $f_{\eta}$, is significantly different between the cold- and hot-seed models. More specifically, in the hot-seed model case $f_{\eta}$ is almost 100$\%$ at all frequency intervals, whereas in the cold-seed model case it stays more or less stable around 50$\%$. In both model cases, the high values of $f_{\eta}$ indicate that a significant amount of the NS luminosity is due to feedback. 

We also found reasonable values (compared to \citealt{Zhang2017}) for the average corona temperature, $kT_e$. In Figure \ref{fig:MCMC_cornerplot} we plot the predicted $kT_e$, for both the cold- and hot-seed model cases, together with the values from \citet{Zhang2017}. The electron temperature, $kT_e$, decreases with QPO frequency in the cold-seed model case, as reported in \citet{Ribeiro2017}, but has a more complex behaviour in the hot-seed model case. However, as we explain later, we did not expect a perfect agreement between our fitted $kT_e$ values and the values presented in \citet{Zhang2017}, given that our time averaged spectral model is necessarily different from theirs, mainly because we use a simple blackbody as the seed photon source. Nevertheless, our predicted values of $kT_e$ for the cold-seed model case agree at the 1$\sigma$ level with the observed values from \citet{Zhang2017}, contrary to the hot-seed case. The temperature of the NS, $kT_s$,  decreases from around 1.8 keV at 570 Hz, to 1 keV at 610 Hz and then remains quite stable in the hot-seed model case, while in the cold-seed model case $kT_s$ is around 0.3 keV at all QPO frequencies.

\begin{figure}
	\includegraphics[width=\columnwidth]{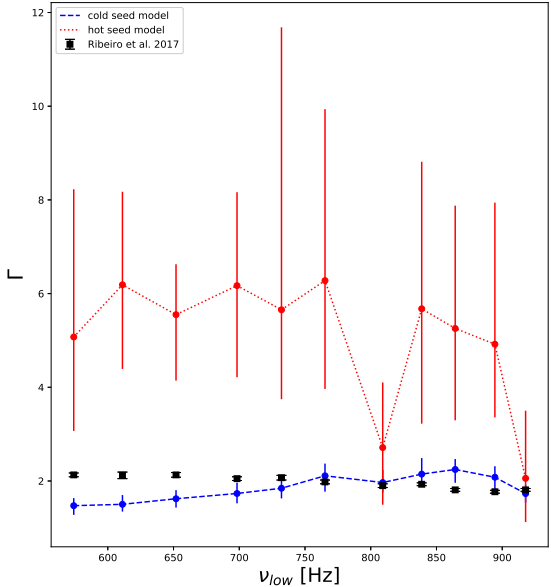}
    \caption{Observed (black squares, \citet{Ribeiro2017}) values of the photon power law index, $\Gamma$, as a function of the frequency of the lower kHz QPO for 4U $1636-53$, and the predicted values based on the hot- (red dotted line) and cold-seed (blue dashed line) model cases. }
    \label{fig:Gammas}
\end{figure}

Finally, we find that in the cold-seed model case the optical depth, $\tau_T$, slightly decreases from around 10.6 at 610 Hz, to 8.5 at 860 Hz. These values are consistent with the observed values of \citet{Zhang2017}, but not with the increasing behaviour with QPO frequency reported in the latter paper and also in \citet{Ribeiro2017}. On the other hand, in the hot-seed model case, the values of $\tau_T$ are significantly smaller with no clear behaviour. The photon power-law index, $\Gamma$, for the hot-seed model case is found to be above 5, on average. For the cold-seed case model, $\Gamma$ increases from around 1.5 at 570 Hz to around 2.2 at 860 Hz and then decreased to 1.7 at 920 Hz. In Figure \ref{fig:Gammas}  we plot the observed $\Gamma$ values from \citet{Zhang2017} and \citet{Ribeiro2017} together with the predicted values of $\Gamma$, given the best-fitting values of $kT_e$ $\tau_T$ and their MCMC uncertainties. The values of $\Gamma$, alongside their propagated uncertainties, are given in Table \ref{tab:retrieve}.

\begin{table*}
\caption{Best-fitting values of the model parameters for the rms and time lag energy spectra of the lower kHz QPO in 4U $1636-53$ as a function of QPO frequency. At each line of the table, the first row corresponds to the best-fitting cold-seed model and the second one to the best-fitting hot-seed model case. }
\label{tab:fit}
\renewcommand{\arraystretch}{2}
\begin{tabular}{cccccccccc}
\hline
   $\nu_{low}$ [Hz] & $L$ [km] & $|\delta \dot{H}_{ext}|$ & $f_{\eta}$& $kT_e$ [keV] & $kT_s$ [keV] & $\tau_T$ & $ \chi^2/d.o.f.$ \\
\hline
574.1 & 3.9$^{+2.2}_{-2.3}$ & 0.08 $\pm \ 0.04$ & 0.7 $\pm \ 0.2$ & 6.7 $\pm \ 1.7$ & 0.4 $\pm \ 0.45$ & 10.4$^{+2.5}_{-2}$ & 14.2/6 
%\vspace{3.8} 
\\ 
& 2 $\pm \ 0.4$ & 0.07$\pm 0.2$ & 0.96 $\pm \ 0.03$ & 8.9 $^{+6.2}_{-4.4}$ & 1.8 $^{+1}_{-0.6}$ & 1.4$^{+1.2}_{-0.6}$ & 11.5/6
\\
\hline
611.2 & 4.7$\pm \ 2.4$ & 0.1 $\pm \ 0.05$ & 0.5 $\pm \ 0.2$ & 5.5$^{+1.4}_{-2.1}$ & 0.3 $^{+0.2}_{-0.06}$ &
10.6$^{+1}_{-1.5}$ & 28.2/6
% \vspace{3.8}
 \\
& 1.3$^{+1.2}_{-0.6}$ &
0.064$^{+0.007}_{-0.02}$ & 
0.97 $\pm \ 0.04$ &
7.9$^{+1.71}_{-2.2}$ &
1$\pm \ 0.1$ &
1.1$\pm \ 0.5$ & 16.5/6 \\ 
\hline
651.9 & 5.3$\pm \ 2.2$ &
0.14 $\pm \ 0.05$ &
0.5 $\pm \ 0.08$ &
4.9 $\pm \ 1.5$ &
0.3 $^{+0.05}_{-0.2}$ &
10.4$\pm \ 1$ & 9.8/8 
%\vspace{3.8}
 \\
& 1.1$\pm \ 0.3$ &
0.08 $\pm \ 0.02$ &
0.97 $\pm \ 0.02$ &
5.8$\pm 1.5$ & 
1 $\pm \ 0.3$ &
1.6$^{+0.3}_{-0.5}$ & 14.8/8
\\
\hline
698.3 & 6.4$^{+3}_{-2.7}$ &
0.12$\pm \ 0.06$ &
0.5 $\pm \ 0.1$ &
4.2 $\pm \ 1.4$ &
0.3$^{+0.02}_{-0.04}$ &
10$\pm \ 1$ & 22.6/8
%\vspace{3.8}
 \\
& 1.1 $\pm \ 0.3$ &
0.08 $\pm \ 0.008$ &
0.97 $\pm \ 0.02$ &
7.4$^{+0.9}_{-1.6}$ &
1 $\pm \ 0.2$ &
1.2 $\pm \ 0.5$ & 10.2/8
\\
\hline
732.1 & 7.5 $\pm \ 2.6$ & 0.13 $\pm \ 0.06$ &
0.5 $\pm \ 0.1$ &
3.6 $\pm \ 1.1$ &
0.3 $\pm \ 0.03$ &
10 $\pm \ 1$ & 15.1/7
%\vspace{3.8}
 \\
& 1.6 $\pm \ 0.6$ &
0.1 $\pm \ 0.04$ &
0.96 $\pm \ 0.03$ &
8.7$^{+6.7}_{-4.3}$ &
1.5 $\pm \ 0.4$ &
1.2 $^{+1.8}_{-0.4}$ & 16.7/7
\\
\hline
765.2 & 8.7 $\pm \ 2.8$ &
0.094 $\pm \ 0.04$ &
0.6 $\pm \ 0.2$ & 
2.7 $\pm \ 1$ &
0.3$^{+0.04}_{-0.1}$ &
9.8 $\pm \ 1$ & 27.9/8
%\vspace{3.8}
 \\
& 1.2 $\pm \ 0.5$ &
0.1 $\pm \ 0.04$ &
0.96$^{+0.04}_{-0.08}$ &
5.8 $\pm \ 2.6$ &
1.2 $\pm \ 0.3$ &
1.4$^{+1.1}_{-0.6}$ & 31.4/8
\\
\hline
809.1 & 5.4 $\pm \ 1.6$ &
0.09 $\pm \ 0.03$ & 
0.5 $\pm \ 0.09$ &
2.9 $\pm \ 0.9$ &
0.31 $\pm \ 0.03$ &
10.3 $\pm \ 0.8$ & 32.5/8
%\vspace{3.8}
 \\
& 1.3$^{+0.3}_{-0.4}$ &
0.13 $\pm \ 0.03$ &
0.8 $\pm \ 0.09$ &
4.9$^{+1.9}_{-1.6}$ &
1 $\pm \ 0.2$ &
4.9 $\pm \ 3$ & 55/8
\\
\hline
838.8 & 7.4 $\pm \ 2$ &
0.08 $\pm \ 0.03$ &
0.6 $\pm \ 0.1$ &
2.8$^{+0.9}_{-0.6}$ &
0.3$^{+0.02}_{-0.06}$ &
9.4 $\pm \ 1$ & 29.2/8
%\vspace{3.8}
 \\
& 1.4 $\pm \ 0.4$ &
0.1 $\pm \ 0.03$ &
0.95 $\pm \ 0.04$ &
6.6$^{+3}_{-3.3}$ &
1.3$^{+0.3}_{-0.2}$ &
1.45$^{+1}_{-0.7}$ & 31.7/8
\\
\hline
864.3 & 5 $\pm \ 1$ &
0.06 $\pm \ 0.02$ &
0.7 $\pm \ 0.08$ &
2.9 $\pm \ 0.7$ &
0.25$^{+0.06}_{-0.02}$ &
8.5 $\pm \ 0.7$ & 70/7
%\vspace{3.8}
 \\
& 1.1$^{+0.6}_{-0.4}$ &
0.07 $\pm \ 0.02$ &
0.98$^{+0.01}_{-0.03}$ &
6.5$^{+4.4}_{-2.7}$ &
1 $\pm \ 0.2$ & 6.1
1.6$^{+1}_{-0.7}$ & 41/7
\\
\hline
894.4 & 5.2 $\pm \ 1.6$ &
0.005 $\pm \ 0.003$ &
0.58$^{+0.1}_{-0.08}$ &
2.7$^{+0.5}_{-1}$ &
0.3 $\pm \ 0.03$ &
9.9 $\pm \ 1$ & 53.8/8
%\vspace{3.8}
 \\
& 1.2 $\pm \ 0.4$ &
0.074$^{+0.03}_{-0.01}$ &
0.94 $\pm \ 0.04$ &
5.5$^{+2.5}_{-1.7}$ &
1 $\pm \ 0.2$ &
2$^{+1.7}_{-0.8}$ & 52.8/8
\\
\hline
917.8 & 5.1$^{+1.9}_{-1.6}$ &
0.055 $\pm \ 0.03$ &
0.66 $\pm \ 0.08$ &
5$^{+1.2}_{-1.4}$ &
0.25 $\pm \ 0.07$ &
9.2 $\pm \ 0.9$ & 55.5/8
%\vspace{3.8}
 \\
& 1.5 $\pm \ 0.5$ &
0.048$^{+0.03}_{-0.01}$ &
0.97$^{+0.02}_{-0.09}$ &
6.7$^{+1.8}_{-1.9}$ &
1.3 $\pm \ 0.2$ &5.9$^{+6}_{-3.8}$ & 78.8/8
\\
\hline
\end{tabular}
\end{table*}

\subsection{The leading oscillation}

\begin{figure*}
	\includegraphics[width=15cm,height=17cm]{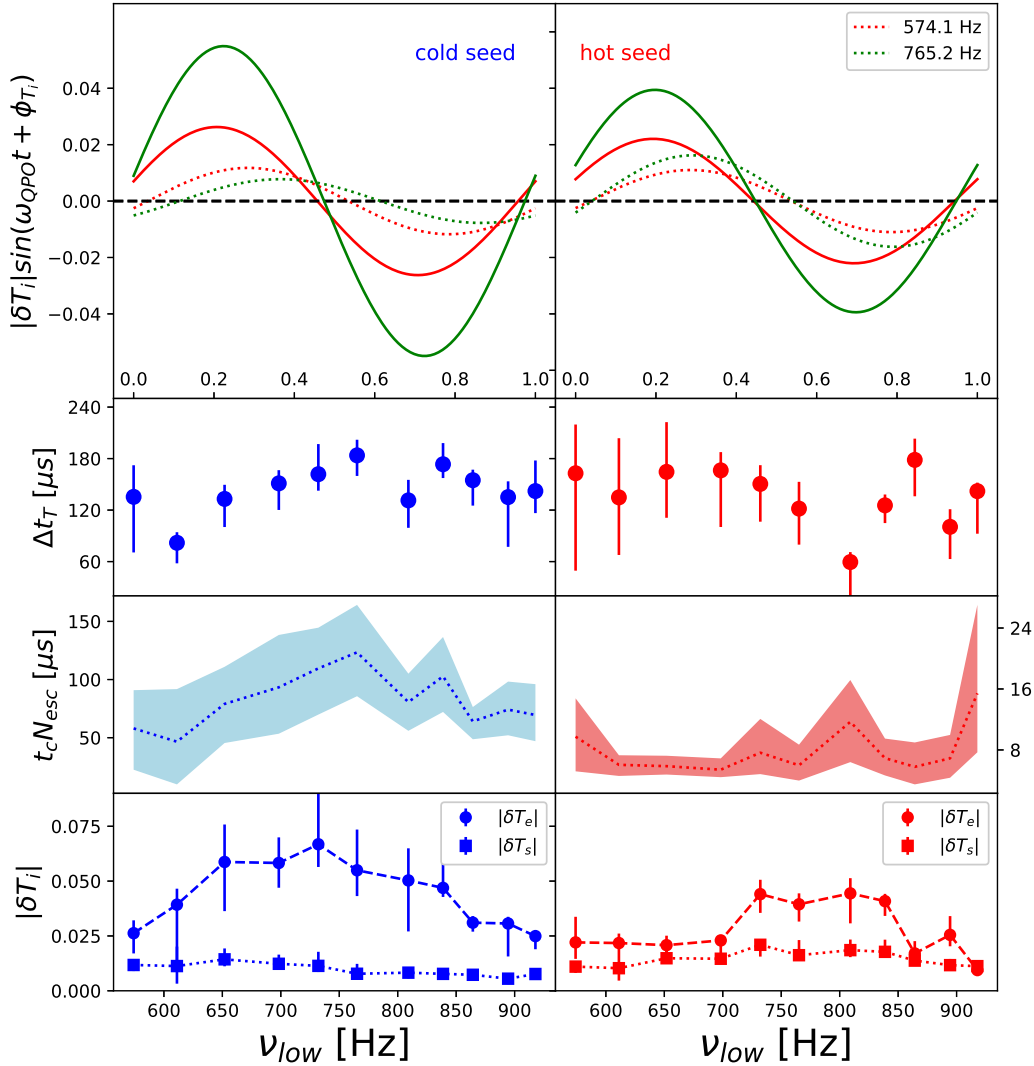}
    \caption{Comparison between the oscillating temperature and oscillation amplitude of the corona, $T_e$, and NS, $T_s$, in 4U $1636-53$ as a function of the frequency of the lower kHz QPO for the cold- (left column) and hot-seed model (right column). In the top panel, the solid lines represent the mock oscillations of the corona temperature, $T_e$, while the dotted lines represent the mock oscillations of the NS temperature, $T_s$, at two arbitrary lower kHz QPO frequencies, denoted with different colors. In the second panel, we plot the time lag between $ T_e$ and $T_s$ as a function of the QPO frequency. In the third panel, we plot the average photon escape time, $t_c N_{esc}$, as a function of QPO frequency. In the bottom panel, we compare the oscillation amplitudes $|\delta T_e|$ and $|\delta T_s|$. We must note that the quantities plotted here are not directly observed but retrieved from the model, after fitting the rms and time lag data.} 
    \label{fig:mock_model}
\end{figure*}

Our solving scheme, described in detail in section \ref{sec:model} and Appendix \ref{ap:pert},
solves for the complex amplitudes $\delta n_{\gamma}(E)$, $\delta T_e$ and $\delta T_s$ self consistently. Since no mathematical information about which one of the two oscillations leads in time is put in equation (\ref{eq:CODE}), a priori, the solution will dictate which of $\delta n_{\gamma}$, $\delta T_e$ or $\delta T_s$ will be the leading signal at the time of escape from the corona. At all QPO frequencies, we find that the oscillation of the NS temperature, $\delta T_s$, lags the oscillation of the corona temperature, $\delta T_e$, by  $\sim 150 \ \mu s$ on average, both for the cold- and hot-seed model. In order to visualize our result, we created a mock model for the temperatures of the corona and NS, both for the cold- and hot-seed model cases. We represent both of the temperatures $T_e$ and $T_s$ as simple sinusoidal signals with amplitudes $|\delta T_e|$ and $|\delta T_s|$, and phases $\phi_{T_e}$ and $\phi_{T_s}$, respectively, normalized around zero. All the quantities mentioned above are taken directly from our best-fitting cold- and hot-seed model at the given frequency interval. Our wave-like mock models are presented in Figure \ref{fig:mock_model} for three different QPO frequencies, alongside the predicted time lags, $\Delta t_{T}$, between the leading and lagging oscillations for the same QPO frequencies. We plot the time lag, $\Delta t_{T}=t_{T_e}-t_{T_s}$, between the response of the corona and the NS, in the second panel of Figure \ref{fig:mock_model}. We summarise the retrieved values of $\Delta t_{T}$ in Table \ref{tab:retrieve}, for each QPO frequency.

Based on our model, we can explain both why the oscillating NS temperature leads in time, and the dependence of the time delay between the two oscillations, $\Delta t_T$,  upon QPO frequency. On the first question, the high feedback fraction that we find in all our fits means that a significant fraction of the emitted soft photons, from the NS surface, is delayed because of the feedback mechanism. If the temperature of the NS is not oscillating independently, then the only allowed oscillation of the NS temperature would be the one caused by the feedback photons, whose number density oscillates at the QPO frequency. The time that it takes for a photon to impinge back onto the NS surface is hard to calculate using random walk arguments. However, we can assume that it is of the order of $t_c N_{esc}$, which is the average escape time from the corona. Thus, by the time the QPO is produced, the NS temperature, through feedback, responds after $\sim t_c N{esc}$. If the QPO is produced by an oscillation of the corona temperature, then $\delta T_e$ would naturally lead in time. However, even if the corona responds to an already existing QPO ($\delta n_{\gamma}$), through Compton cooling, the extremely short, compared to $t_c N_{esc}$ $\sim 10^{-5} \ s$ , electron-electron collision characteristic timescale ($\sim 10^ {-10} \ s$) would make the corona temperature to respond almost instantaneously to the presence of an oscillating photon number density inside it.

From the above reasoning, one would expect $\Delta t_T$ to be of the order of $t_c N_{esc}$. That brings us to the second question, about the measured time delay $\Delta t_T$. This is well correlated (in the cold-seed model case) to the time $t_cN_{esc}$, needed for photons, on average, to escape the medium. We plot $\Delta t_T$ and $t_c N_{esc}$ as a function of QPO frequency in the second and third panels of Figure \ref{fig:mock_model}, respectively. This agreement between the time difference, $\Delta t_T$, and the factor $t_c N_{esc}$, predicted by random walk arguments, was expected and therefore it serves as a sanity check of our model.

\subsection{Thermodynamic properties of the corona}

\begin{figure}
	\includegraphics[width=\columnwidth]{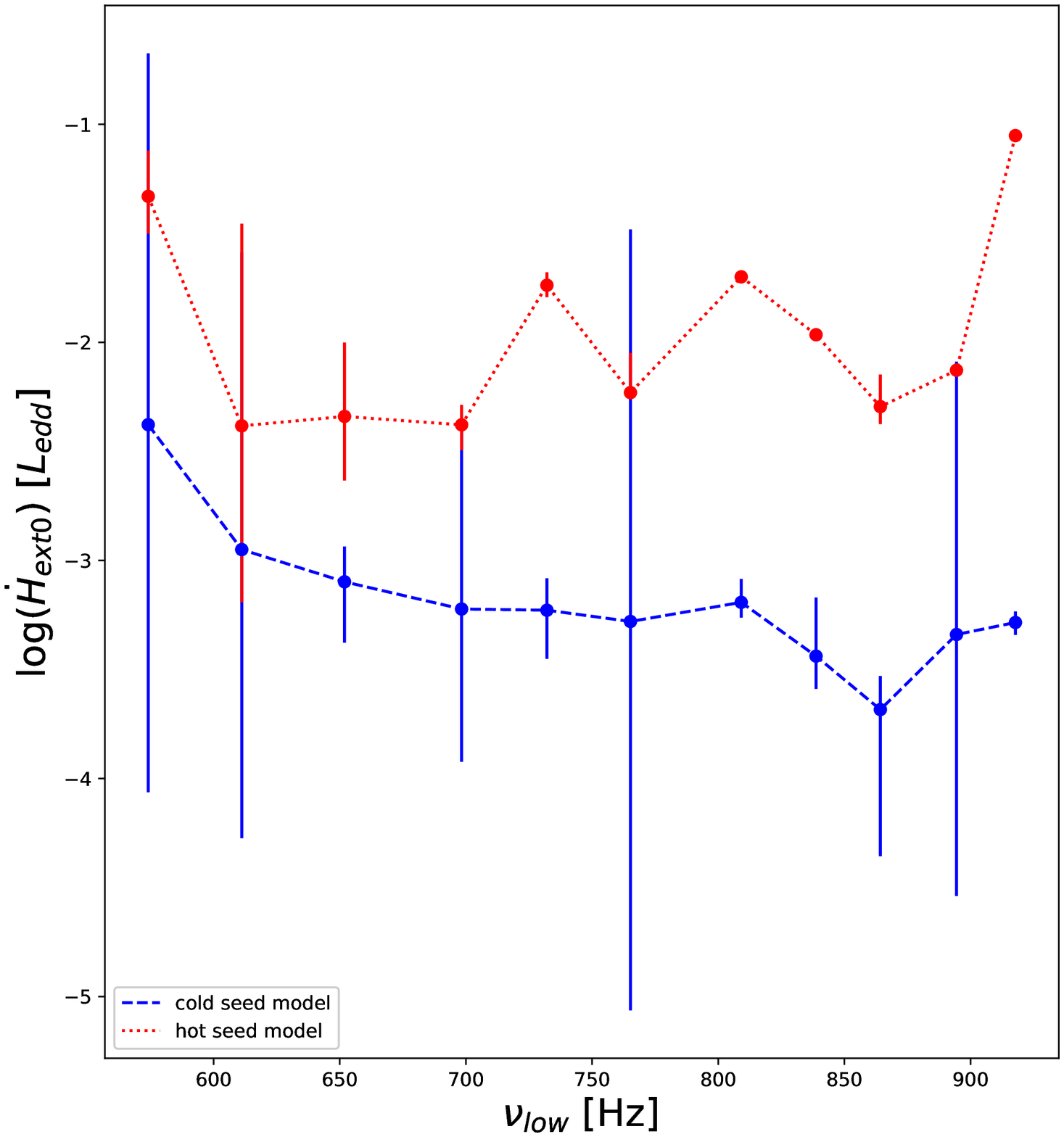}
    \caption{The retrieved average external heating rate, $\dot{H}_{ext0}$, for 4U $1636-53$ in  units of the Eddington luminosity, as a function of the frequency of the lower kHz QPO for the cold- (blue dashed line) and hot-seed (red dotted line) model cases. }
    \label{fig:Hext}
\end{figure}

In our model we can retrieve the average external heating that the corona requires per unit time, per scattering and per electron, $\dot{H}_{ext0}$, in order to maintain equilibrium. This external heating rate can be calculated from equation \ref{eq:Hext} and the dependence upon energy, through the Kein-Nishina cross section, simply means that the cooling is less efficient at higher energies (i.e. less external heating is required). Based on our best-fitting parameters, we can estimate that $\dot{H}_{ext0}$ is, on average, below $2 \%$ of the Eddington luminosity both for the cold- and hot-seed model. In the hot-seed model case, we find that $\dot{H}_{ext0}$ is systematically larger than in the cold-seed case by almost an order of magnitude. We plot the dependence of $\dot{H}_{ext0}$ to QPO frequency, for the cold- and hot-seed model case, in Figure \ref{fig:Hext}. In the cold-seed case, $\dot{H}_{ext0}$ is on average much smaller than the observed luminosity of 4U $1636-53$, which is around 0.1 Eddington, at the state where the kHz QPOs are observed. The reason that we find low values of $\dot{H}_{ext0}$ is because we only study the kHz QPOs which have time lags of the order of micro seconds, which in turn means that the corona sizes are expected to be small as we find here. In other words, we only study the inner part of the corona while the full corona could be more extended, thus requiring a much higher heating rate. 

\begin{figure}
	\includegraphics[width=\columnwidth]{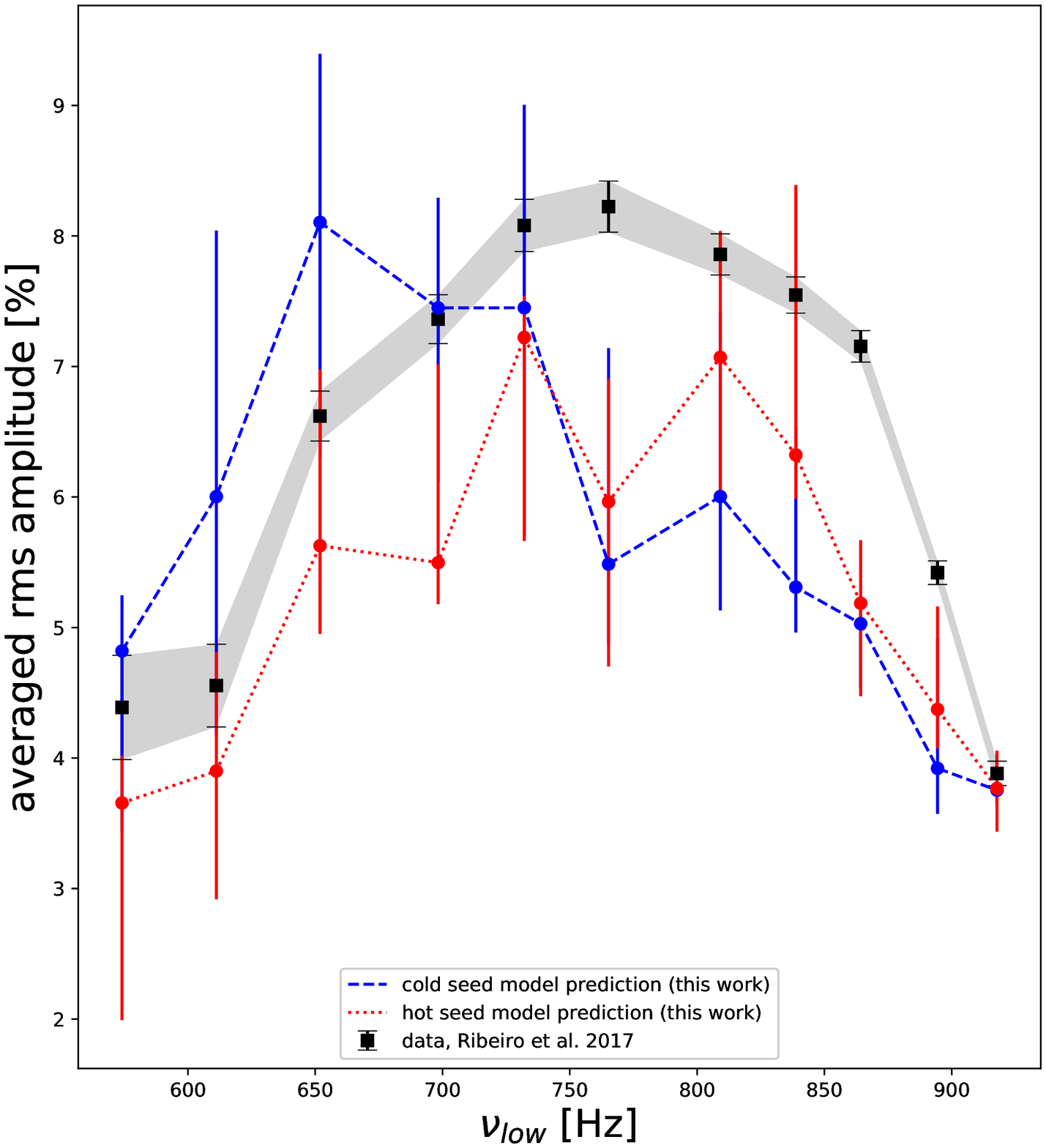}
    \caption{The energy averaged fractional rms amplitude measurements as a function of the frequency of the lower kHz QPO for 4U $1636-53$ (\citealt{Ribeiro2017}; grey squares and shaded area), alongside our cold- and hot-seed model predictions with blue dashed and red dotted lines, respectively.}
    \label{fig:avg_rms}
\end{figure}

\begin{figure}
	\includegraphics[width=\columnwidth]{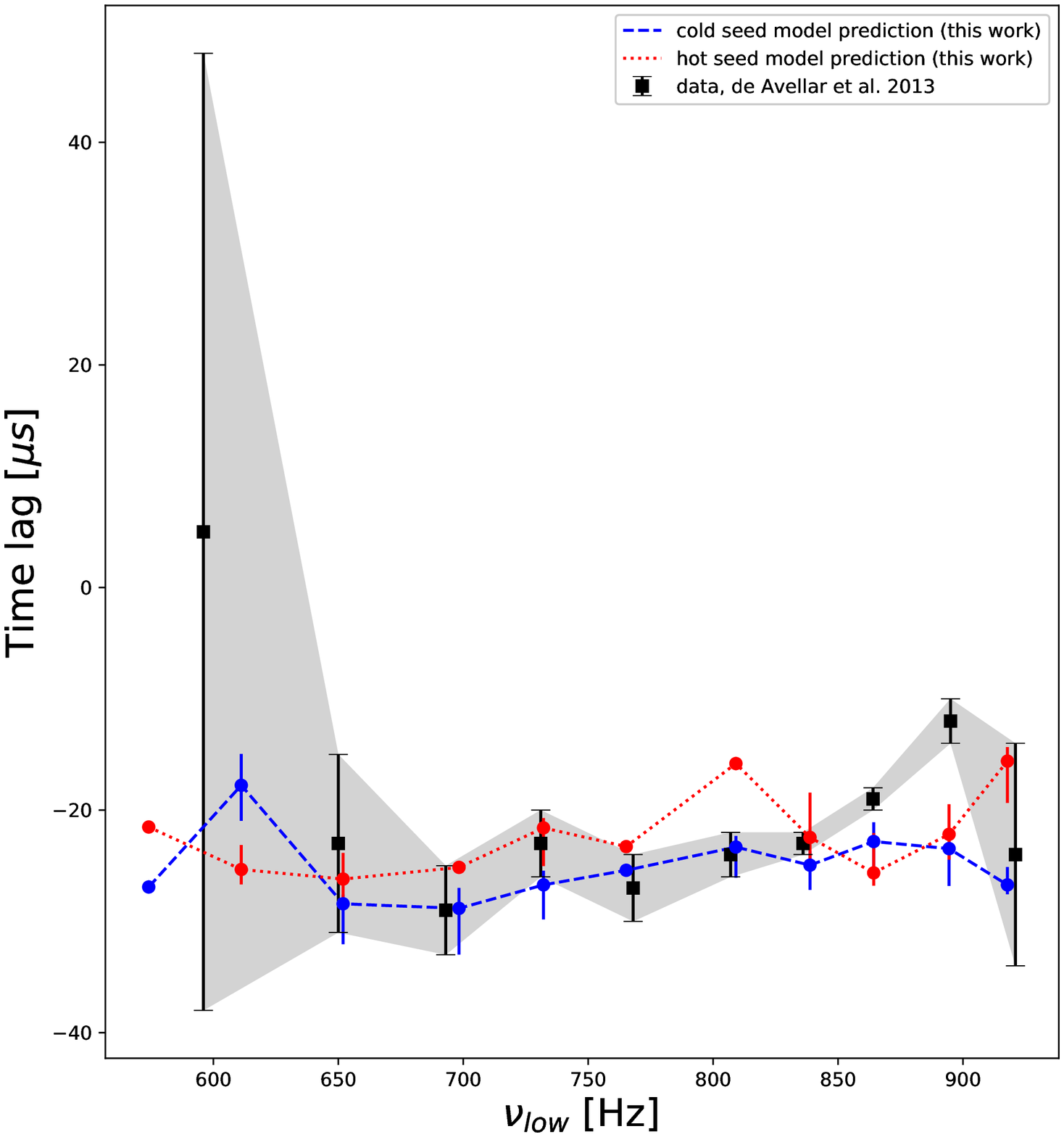}
    \caption{The energy averaged time lags of the lower kHz QPO in 4U $1636-53$ vs QPO frequency (\citealt{Avellar2013}; grey squares and shaded area), alongside the cold- and hot-seed model predictions with blue dashed and red dotted lines, respectively.}
    \label{fig:avg_lag}
\end{figure}

On top of the heating rate, $\dot{H}_{ext0}$, an oscillation of this heating rate, $\delta \dot{H}_{ext}$, is used to explain the observations. Since $\delta \dot{H}_{ext}$ is a free parameter in our model, we can assume that its complex phase is zero, since it does not affect the phase (or time) delay between $\delta T_e$ and $\delta T_s$ or the QPO time lags, given that they are calculated with respect to a reference energy band. Given the fact that $|\delta \dot{H}_{ext}|$ plays the role of a normalization for the modulus of the QPO amplitude, |$\delta n_{\gamma}$|, it is not surprising that $|\delta \dot{H}_{ext}|$ is correlated to the energy averaged rms (Figure \ref{fig:avg_rms}), since the latter may be thought of us approximately the area under the curve of rms as a function of energy. A direct comparison between our best-fitting cold- and hot-seed model and the energy averaged rms and time lags presented in \citet{Ribeiro2017} and \citet{Avellar2013} respectively is shown in Figures \ref{fig:avg_rms} and \ref{fig:avg_lag}. We calculated the energy averaged rms in the full energy band (nominally 2-60 keV). For the comparison to the time lags of \citet{Avellar2013}, we computed the time lag of the 12-20 keV band with respect to the 4-12 keV band using the best-fitting parameters in each frequency interval.

 Finally, the NS temperature oscillation amplitude, $|\delta T_s|$, remained constant, within uncertainties, at 1$\%$ both in the cold- and hot-seed model case, at all QPO frequencies. However, the oscillation amplitude of the corona temperature, $|\delta T_e|$, in the cold-seed model case increased from $\sim$2.6$\%$ at 574 Hz to around 7$\%$ at 732 Hz, and then decreased again to 2.5$\%$ at 918 Hz. In the hot-seed model case, $|\delta T_e|$ was almost as low as the NS temperature oscillation amplitude, $|\delta T_s|$, at all frequencies albeit a small range around 809 Hz were a maximum of $4\%$ was found. In the bottom panel of Figure \ref{fig:mock_model} we plot these amplitudes as a function of QPO frequency. We present the exact values of the retrieved parameters $|\delta T_e|$, $|\delta T_s|$ and $\dot{H}_{ext0}$ in Table \ref{tab:retrieve}.

\begin{table*}
\caption{Physical parameters retrieved from our model after the fitting to the lower kHz QPO in 4U $1636-53$, at each QPO frequency interval. Each parameter is accompanied by the 1$\sigma$ uncertainty around the best fitting, median, value. }
\label{tab:retrieve}
\def\arraystretch{1.5}
\renewcommand{\arraystretch}{2}

\begin{tabular}{ccccccccc}
\hline
   $\nu_{low}$ [Hz] & $|\delta T_e|$ & $|\delta T_s|$ & $\Delta t_T$ [$\mu s$]& $\log{\dot{H}_{ext0}}$ [$L_{edd}$]& $\eta$ & $\Gamma$    \\
\hline
574.1 & 0.026$^{+0.005}_{-0.009}$ & 0.011 $\pm \ 0.003$ & 135$^{+40}_{-70}$ & 
-2.3 $\pm \ 1.7$ &
0.1$^{+0.06}_{-0.2}$ &
1.5 $\pm \ 0.2$
%\vspace{3.8}
 \\
& 0.022 $\pm \ 0.01$ &
0.011 $\pm \ 0.003$ &
162$^{+57}_{-110}$  &
-1.3 $\pm \ 0.2$ &
0.85$^{+0.07}_{-0.2}$ &
5$^{+3.1}_{-2}$
\\
\hline
611.2 & 0.042$^{+0.007}_{-0.03}$ & 0.012 $\pm \ 0.009$ &
128$^{+12}_{-24}$&
-2.8 $\pm \ 1.3$ &
0.1$^{+0.1}_{-0.02}$ &
1.55 $\pm \ 0.2$
%\vspace{3.8}
 \\
& 0.021$^{+0.004}_{-0.01}$ &
0.01$^{+0.001}_{-0.006}$ &
135 $\pm \ 68$ &
-2.3 $\pm \ 0.9$ &
0.88$^{+0.02}_{-0.2}$ &
6.2 $\pm \ 2$ 
\\
\hline
651.9 & 0.058$^{+0.01}_{-0.02}$ &
0.014$^{+0.005}_{-0.003}$ &
133$^{+16}_{-33}$ &
-3$^{+0.2}_{-0.3}$ &
0.1$^{+0.08}_{-0.02}$&
1.6 $\pm \ 0.2$
%\vspace{3.8}
 \\
& 0.02$^{+0.004}_{-0.008}$ &
0.014$^{+0.005}_{-0.002}$ &
165$^{+58}_{-54}$ &
-2.3 $\pm 0.3$ &
0.88$^{+0.07}_{-0.05}$ &
5.5$^{+1}_{-1.4}$
\\
\hline
698.3 & 0.058 $\pm \ 0.01$ &
0.012$^{+0.004}_{-0.003}$ &
151$^{+15}_{-31}$&
-3.2$^{+0.9}_{-0.7}$ &
0.14$^{+0.08}_{-0.03}$ &
1.7 $\pm \ 0.2$
%\vspace{3.8}
 \\
& 0.022$^{+0.003}_{-0.009}$ &
0.014 $\pm \ 0.002$ &
166$^{+21}_{-66}$ &
-2.3 $\pm \ 0.1$ &
0.89 $\pm \ 0.06$ &
6.2 $\pm \ 2$
\\
\hline
732.1 & 0.066$^{+0.03}_{-0.01}$ &
0.011$^{+0.006}_{-0.002}$ &
162$^{+35}_{-19}$&
-3.2 $\pm \ 0.2$ &
0.17$^{+0.1}_{-0.03}$ &
1.8 $\pm \ 0.2$
%\vspace{3.8}
 \\
& 0.044$^{+0.007}_{-0.009}$ &
0.02$^{+0.002}_{-0.005}$ &
150$^{+22}_{-44}$&
-1.7 $\pm \ 0.06$ &
0.87$^{+0.08}_{-0.02}$ &
5.6$^{+6}_{-2}$
\\
\hline
765.2 & 0.054$^{+0.02}_{-0.01}$ & 
0.008$^{+0.005}_{-0.001}$ &
184$^{+18}_{-24}$ &
-3.2 $\pm \  1.8$ &
0.27$^{+0.03}_{-0.1}$ &
2.1 $\pm \ 0.3$ &
%\vspace{3.8}
 \\
& 0.04$^{+0.005}_{-0.008}$ &
0.016$^{+0.007}_{-0.002}$ &
122$^{+31}_{-42}$ &
-2.2$^{+0.2}_{-0.02}$ &
0.89$^{+0.07}_{-0.5}$ &
6.3$^{+3.6}_{-2.3}$ 
\\
\hline
809.1 & 0.05 $\pm \  0.02$ &
0.008 $\pm \ 0.002$ &
131$^{+24}_{-32}$ &
-3.19 $\pm \ 0.1$ &
0.21 $\pm \ 0.1$ &
1.9 $\pm \ 0.3$
%\vspace{3.8}
 \\
& 0.04$^{+0.007}_{-0.01}$ & 
0.018 $\pm \ 0.005$ &
59$^{+12}_{-50}$ &
-1.7 $\pm \ 0.005$ &
0.55$^{+0.05}_{-0.1}$ &
2.7 $\pm \ 1.4$
\\
\hline
838.8 & 0.046$^{+0.02}_{-0.004}$ &
0.008$^{+0.001}_{-0.003}$ &
173$^{+25}_{-16}$ &
-3.4$^{+0.3}_{-0.1}$ &
0.3$^{+0.03}_{-0.08}$ &
2.14 $\pm \ 0.3$ 
%\vspace{3.8}
 \\
& 0.04$^{+0.003}_{-0.007}$ &
0.018$^{+0.006}_{-0.001}$ &
126$^{+13}_{-21}$ &
-1.9 $\pm \ 0.01$ &
0.87$^{+0.05}_{-0.2}$ &
5.6$^{+3.1}_{-2.4}$
\\
\hline
864.3 & 0.031$^{+0.003}_{-0.004}$ &
0.007$^{+0.0005}_{-0.002}$ &
155$^{+12}_{-30}$ &
-3.7$^{+0.2}_{-0.7}$ &
0.33$^{+0.2}_{-0.03}$ &
2.2 $\pm \ 0.3$
%\vspace{3.8}
 \\
& 0.017$^{+0.006}_{-0.002}$ &
0.013 $\pm \ 0.002$ &
178$^{+25}_{-42}$ &
-2.3$^{+0.2}_{-0.08}$ &
0.88$^{+0.02}_{-0.2}$ &
 5.2$^{+2.6}_{-1.9}$
\\
\hline
894.4 & 0.03$^{+0.003}_{-0.02}$ &
0.006$^{+0.0008}_{-0.002}$ &
135$^{+18}_{-58}$ &
-3.3 $\pm \ 1.2$ &
0.26$^{+0.07}_{-0.1}$ &
 2 $\pm \ 0.3$
%\vspace{3.8}
 \\
& 0.025$^{+0.009}_{-0.005}$ &
0.01$^{+0.003}_{-0.001}$ &
100$^{+21}_{-38}$ &
-2.1 $\pm \ 0.01$ &
0.83$^{+0.03}_{-0.1}$ &
4.9$^{+3}_{-1.5}$ 
\\
\hline
917.8 & 0.025$^{+0.001}_{-0.006}$ &
0.0076 $\pm \ 0.0007$ &
142$^{+36}_{-26}$ &
-3.2 $\pm \ 0.06$ &
 0.16$^{+0.1}_{-0.09}$ &
1.7 $\pm \ 0.2$ 
%\vspace{3.8}
 \\
& 0.0095 $\pm \ 0.0002$ &
0.01$^{+0.002}_{-0.0007}$ &
142$^{+20}_{-50}$ & 
-1 $\pm \ 0.02$ &
0.51$^{+0.2}_{-0.1}$ &
 2$^{+1.5}_{-0.9}$
\\
\hline
\end{tabular}
\end{table*}

\section{Discussion}

We fitted the rms and time lag spectra of the lower kHz QPO of the NS LMXB 4U $1636-53$ at different QPO frequencies with a physical model for the variability. Our model assumes that soft photons from the NS surface only, are up-scattered inside a hot and homogeneous corona that is modeled as a spherically symmetric shell that extends to a certain distance, $L$, from the NS surface. In our simplified model the accretion disc is technically assumed to lie outside of the corona. Furthermore, our corona is not associated with the boundary layer, although the latter might in reality be included in what we call the Comptonising medium. Our model takes into account the fact that after an average number of scatterings, a fraction of the hard photons, $\eta$, that have not yet escaped the corona will return to the NS surface, forming a feedback loop, and will cause a delayed heating of the NS up to a higher observed temperature, $T_s$. Our fits yield two statistically acceptable solutions, one with a cooler (cold-seed) and another with a hotter (hot-seed) NS surface. Through our fits, we can place a lower limit to the size of the corona, $L$, between 1-8 km, in the sense that this is the required corona size in order to explain the lower kHz QPO properties based on inverse Compton scattering. Our fits indicate that the fraction of the NS luminosity that is due to feedback from the corona, $f_{\eta}$, is more than 50$\%$ in all model cases. We also find that the oscillations of the NS temperature lag the oscillations of the corona temperature by 150 $\mu s$ on average, due to the finite travel time of the feedback photons from the corona back to the NS surface. Thus, the soft lags, as explained by our model, are due to the finite escape time in the corona of photons illuminating the neutron star surface. The oscillating thermodynamic properties of the corona, namely the external heating rate, $\dot{H}_{ext}$, and corona temperature, $T_e$, exhibit a maximum variability when the lower kHz QPO frequency is at 700 Hz, suggesting a resonance between the source of soft photons (in our case the NS surface) and the corona. To address the cold- and hot-seed model degeneracy that we encountered, we note that in the cold-seed case the inferred values of the photon power-law index, $\Gamma$, tend to be more compatible with the measured values of \citet{Zhang2017}. Furthermore, the inferred time delay between the corona and NS temperature oscillations are in good agreement with the photon diffusion time in the corona, deduced by random walk arguments, only in the cold-seed model case. In general, the corona appears to be the more active among the two components, corona and NS, and our results suggest that the corona is primarily responsible for the radiative properties of the lower kHz QPOs in LMXBs.   

\subsection{Insight about the properties of the corona}

We find a mean (over the different frequency intervals) electron density of $\bar{n}_e \sim 10^{25} \ m^{-3}$ for both the cold- and hot-seed model case, and an average electron temperature, $kT_e$, of around 4 keV and 6.8 keV for the cold- and hot-seed case, respectively. The resulting Debye length, $\lambda_D$, is $\sim 8 \times 10^{-3}m$. The size of the corona that we find here is significantly larger than this, i.e. $\lambda_D << L$, and therefore the corona can be treated as nearly charge neutral, which allows us to assume that the number density of ions, $n_i$, is equal to $n_e$. We also estimate a Coulomb mean free path of about $ 7 \times 10^{-2}$ m for the electron-ion scattering and so the average electron temperature, $T_e$, can be taken as equal to the average ion temperature, $T_i$, assuming thermal equilibrium has been achieved through ion-electron collisions. We must note that since thermalisation between ions and electrons happens at a rate $\frac{m_e}{m_i}$ times slower than the scattering rate, the "thermal equilibration length" will be $\sim 123$ m, which is still small compared to the km wide corona. Finally, an estimate of the outward radiation pressure in the corona, $P_{rad} = 4\sigma T_e^4 / 3c$, based on our estimated electron temperature, yields a value of $ \sim 10^{16}$ dynes cm$^{-2}$, which agrees, to an order of magnitude, with the pressure due to gravity, $P_g = n_e m_p g L$, assuming a typical surface gravity of around $g= 10^{12} - 10^{13}$ m s$^{-2}$ for the NS in 4U $1636-53$. This balance between gravity and radiation pressure emerges without us forcing any condition for hydrostatic equilibrium and thus supports the idea that a spherical corona can indeed be naturally sustained.

\subsection{On the corona geometry and seed photon source type}

As discussed in section \ref{sec:model}, the current version of our model assumes that the NS surface is the source of seed photons for Comptonisation and no extra component is added. However, \citet{Zhang2017} fitted the spectra of NS 4U $1636-53$, during the occurence of the kHz QPOs, using the model $nthcomp$ of $XSPEC$ with a disc blackbody seed source and a simple blackbody on top of that. The timing data that are used in this work come from \citet{Ribeiro2017} and \citet{Ribeiro2019}, the first of which used the best-fitting specral parameters of \citet{Zhang2017}. Thus any comparison between our best-fitting values of $kT_e$, $kT_s$ and $\tau_T$, or their dependence on QPO frequency, to the results of \citet{Ribeiro2017} must take into account the different spectral approaches. In fact, it is physically more accurate for our model to use a factor of $\frac{1}{2}$ instead of $\frac{1}{3}$ in equation (\ref{eq:Nesc}), because the former is correct when the seed source intensity is a delta-like function at the center of the corona, while the latter is correct when the source intensity is distributed sinusoidally from the center of the corona, as was suggested by \citet{Sunyaev1980}. However, since our variability model is constructed on top of an equilibrium solution, we need this equilibrium solution to resemble the observed energy spectrum as much as possible, thus the use of $\frac{1}{3}$. In principle, one should assume that the seed photons could come both from the surface of the NS and part of the inner accretion disc, assuming the corona is sufficiently large.

Another crucial factor when modelling thermal Comptonisation semi-analytically, is the corona geometry. The corona geometry is regulated through the spatial dependence of the scattering optical depth $\tau$ (or $\tau_T$). The simplest possible geometry is a spherical shell and that is what we assume in this work. However, the geometry of the corona of LMXBs is believed to be more complex and for that reason it is very common to use Monte Carlo methods to model thermal Comptonisation, since there is freedom in choosing the desired geometry. In timing analyses, such as the one presented in \citet{Ribeiro2019}, it would be nearly impossible to use Monte Carlo methods to study the timing properties of the QPOs, so semi-analytical models such as the one presented in \citet{Lee1998}, \citet{Lee2001}, \citet{Kumar2014} and this work are extremely useful.

\begin{figure}
	\includegraphics[width=\columnwidth]{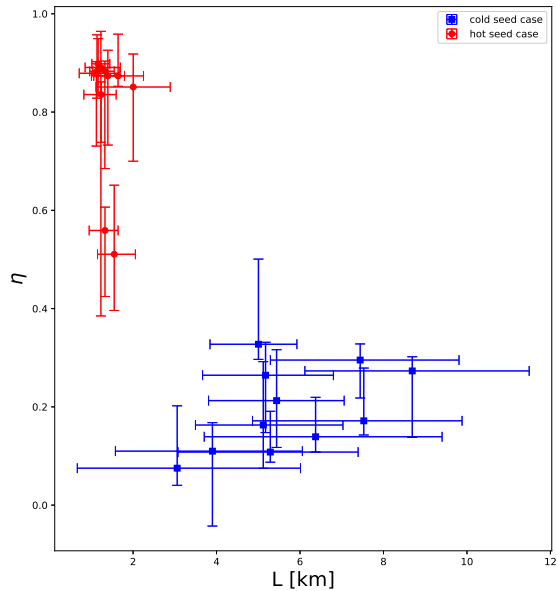}
    \caption{The feedback parameter, $\eta$, retrieved from our model from the fits to the rms and time lag energy spectra of the lower kHz QPO in 4U $1636-53$, plotted against the best-fitting value of the size $L$. The cold- (blue squares) and hot-seed model (red circles) form two separate regimes. }
    \label{fig:eta_vs_L}
\end{figure}

Finally, in Figure \ref{fig:eta_vs_L} we plot the feedback parameter, $\eta$, retrieved for both our cold- and hot-seed model through $f_{\eta}$, against the measured size, $L$, of the corona alongside their MCMC uncertainties. The same quantities were plotted by \citet{Kumar2016} for the system 4U $1608-52$ in the right hand panel of Figure 8 in their paper. \citet{Kumar2016}, however, did not fit the energy dependent rms and time lags at different frequencies but matched the data of \citet{Avellar2013} for the soft time lag between two energy bands with the predictions of a cold- and hot-seed model. In other words, at each lower kHz QPO frequency \citet{Kumar2016} used only one data point for the time lag due to the lack of more data, hence the large uncertainties in their work. Our detailed fits result in significantly smaller uncertainities for $\eta$ and $L$. Thus, a comparison of our data to  Monte Carlo simulations, such as those presented in \citet{Kumar2016MC}, that predicted the feedback, $\eta$, given a specific size, $L$, and an assumed corona geometry, can shed more light on what the most realistic geometry of the corona of NS LMXBs is.

As it is apparent from Figures \ref{fig:fit_flow} and \ref{fig:fit_fhigh} show that our model does not exhibit a break, and a subsequent drop, of the rms amplitude above a certain energy, $E_{break} \sim$12 keV, as reported in \citet{Ribeiro2019}. Mathematically, in order for the rms amplitude to decrease at a certain energy we need to add to our continuum model an extra component, whose flux would exhibit a sufficient incerase at around 12 keV. If the corona is small and optically thick, as we find in this work, the hard X-rays that leave the corona will irradiate the accretion disc; This will increase the disc temperature and produce a reprocessed component that peaks at an energy that depends on the composition of the part of the disc that is irradiated and the irradiating flux itself. Thus, a decrease of the fractional rms as a function of energy is to be expected if we included a reflected component that was less variable in our model. 

Furthermore, since the flux that irradiates the part of the disc that lies outside of the corona is already varying, the reprocessed component will also undergo oscillations. Finally, the inner parts of the disc will be irradiated faster than the outer parts, due to the finite light travel time, which in combination with the already varying incident flux will create a complex time lag as a function of energy. The initial version of the model that we present here can not yet capture all the physical processes that happen around the NS. Although out of the scope of this paper, the next step is a model in which a combination of reprocessed emission from the disc and Comptonisation in the corona can be implemented to improve fitting of the timing properties of kHz QPOs, since reverberation alone can not provide a satisfactory explanation to the energy dependence of the time lags as has been shown by \citet{Cackett2016ApJ} for the NS LMXB 4U $1608-52$.

\subsection{The size of the corona and the rms amplitude of the lower and upper kHz QPOs}

In some models that provide an explanation of the dynamical origin of both the lower and upper kHz QPO, a beat between the NS spin frequency and some other frequency (\citealt{Alpar1985}), e.g., the Keplerian frequency at the marginally stable circular orbit (MSCO), or at the sonic radius of the accretion flow (\citealt{Miller1998}; see also \citealt{Alpar1985}). These models propose that the lower kHz QPO is produced close to the NS surface, and is linked to mass accretion rate instabilities caused by inhomogeneites in the accretion disc. On the other hand, for the upper kHz QPO the flux oscillations could also originate on the NS surface (\citealt{Miller1998}), or they could originate further away and be caused either by occultations of self-luminous disc inhomogeneites or by Doppler beaming of the same inhomogeneites when their radial distance or orbital inclination do not allow occultations (\citealt{Stella1998}). \citet{Gilfanov2003}, proposed that the lower kHz QPO originates in a boundary layer close to the NS surface, and it is associated with Comptonisation, and \citet{Avellar2013} proposed that, regardless of the dynamical mechanism that explains the kHz QPO frequencies, the radiative mechanism that governs the timing properties of the lower and upper kHz QPOs should be different. Recently, \citet{Ribeiro2017} showed that the frequency dependence of the fractional rms amplitude of the upper kHz QPO, although in general different from that of the lower kHz QPO, exhibits the same behaviour as that of the lower kHz QPO around certain QPO frequencies. The latter finding indicates that when the lower and upper kHz QPOs are (separately) observed within a certain QPO frequency range, their variability could be caused by the same radiative mechanism.

From the findings of \citet{Gilfanov2003}, \citet{Avellar2013} and \citet{Ribeiro2017}, we assume that the flux that oscillates at the upper kHz QPO frequency is produced in the disc, the flux that oscillates at the lower kHz QPO frequency is produced close to the NS surface. Furthermore, depending on its size, a spherically symmetric corona can cover both the NS surface and the inner parts of the accretion disc. In this scenario, the balance between the NS radius, $a$, the corona size, $L$, and the radius of the MSCO, $R_{ms}$, determines whether the fractional rms amplitude of the upper kHz QPO is affected by Comptonisation, and thus whether it would resemble the behaviour of the fractional rms amplitude of the lower kHz QPO. \citet{Wang2017} fitted the spectrum of 4U $1636-53$, with different reflection models, and found an inner disc radius, $R_{in}=(5.1-10.3)R_g$, in the intermediate state, during which the lower kHz QPO is typically observed (\citealt{Zhang2017}) ($R_g = GM/c^2$, with $M$ the mass of the neutron star and $G$ and $c$ the gravitational constant and the speed of light, respectively). Assuming a minimum NS mass of 1.55$M_{\odot}$, as suggested by \citet{Bulik2000} based on the maximum kHz QPO frequencies observed in 4U $1636-53$, yields a minimum inner radius of $\sim 12- 23$ km.

Assuming a minimum NS radius of $\sim 11$ km for 4U $1636-53$ (\citealt{Stiele2016ApJ}) which is smaller than the radius of the MSCO, a width of the corona of $\sim 8$ km at 770Hz (see Figure \ref{fig:MCMC_cornerplot}, for the cold-seed source case) would imply that the inner parts of the disc would be covered by the corona at specific frequencies. Based on these arguments, it is very interesting that a hump in the energy averaged rms amplitude of the upper kHz QPO is apparent at 770 Hz (\citealt{Ribeiro2019}), while the energy averaged rms amplitude of the lower kHz QPO peaks at about the same frequency. Although the energy averaged rms amplitude of the upper kHz QPO exhibits a different dependence upon QPO  frequency to that of the lower kHz QPO, \citet{Ribeiro2019} showed that around 770 Hz the energy averaged rms amplitude of the upper kHz QPO resembles that of the lower. If the upper kHz QPO originates in the accretion disc, as its frequency increases and the orbiting material, that causes the QPO, moves inwards there will be a critical frequency of the upper kHz QPO (e.g. 770 Hz) where if the corona is extended enough the orbiting material will cross inside the corona and the upper kHz QPO will be affected by the properties of the corona and thus its radiative mechanism might be the same as that of the lower kHz QPO. We note that the crossing of the orbiting material inside the corona would happen when the lower kHz QPO has a frequency of around 515 Hz given an estimate of the constant separation between the lower and upper kHz QPO in 4U $1636-53$ of about 255 Hz, as reported in \citet{Zhang1997IAUC}.

A more accurate estimation of the critical upper kHz frequency, where the orbiting material crosses inside the corona is non-trivial and would require detailed general relativistic simulations. In particular, the high NS spin of $\sim 582$ Hz reported for 4U $1636-53$ (  \citealt{Strohmayer_2002}) does not allow us to apply corrections to the radius of the MSCO and to the Keplerian frequency of the accreting matter, that are of first order to the dimensionless angular momentum $j\equiv cJ/GM^2$ (\citealt{Hartle1968}; \citealt{MillerLambCook1998} and \citealt{Miller1998}), where $J$ is the orbital angular momentum of the NS. However interesting, such an analysis falls outside the scope of this work.

\subsection{Future prospects with eXTP and NICER}

The enhanced X-ray Timing and Polarimetry mission (eXTP; \citealt{SNZhang2019}), and the Neutron-star Interior Composition ExploreR (NICER; \citealt{Gendreau2012}), are two very promising missions that can help validate some of the results presented here. NICER is currently active on the International Space Station while eXTP is expected to be launched in the mid 2020s. The improved timing and spectral capabilities of both eXTP and NICER will allow phase resolved spectroscopy of LMXBs (see \citealt{Zand2019} for an extensive discussion on the science with eXTP). The variability and time delay between the time dependent temperatures $T_e$ and $T_s$ can be measured and compared to the predictions of the model presented here (e.g Figure \ref{fig:mock_model}). Furthermore, measuring NS radii with NICER through pulse profile modeling (\citealt{Ozel2016ApJ}), combined with knowledge about the size of the corona, can help us understand whether it is correct to use the accretion disc as a source of seed photons in the case of a spherical corona.

\section{Conclusion}

We developed a numerical model that predicts the energy dependence of the rms amplitude and time lags of QPOs in NS LMXBs. Using observations of the lower kHz QPOs in the NS LMXB 4U $1636-53$ we propose that the dependence of the variability of this QPO upon energy and QPO frequency is due to the surrounding Comptonising medium (corona) whose temperature exhibits the same dependence upon QPO frequency. We demonstrated that this model allows one to fit the timing properties of QPOs and get constraints on the thermodynamic properties and spatial extent of the corona. More importantly, our model makes predictions that can be tested with current and future instruments and thus can shed light on whether Comptonisation is the mechanism responsible for the radiative properties of the lower kHz QPOs in NS LMXBs.    

\section*{Acknowledgements}

E.M.R acknowledges the support from Conselho Nacional de Desenvolvimento Cient\`ifico e Tecnol\`ogico (CNPq - Brazil). D.A acknowledges support from the Royal society. O.B. is grateful for the support of a University of Southampton Diamond Jubilee
International Visiting Fellowship. This work is part of the research programme Athena with project
number184.034.002, which is (partly) financed by the Dutch Research
Council (NWO). This research has made use of data obtained from the High Energy Astrophysics Science Archive Research Center, provided by NASA's Goddard Space Flight Center.

%%%%%%%%%%%%%%%%%%%%%%%%%%%%%%%%%%%%%%%%%%%%%%%%%%



\bibliographystyle{mnras}
\bibliography{mnras_template} % if your bibtex file is called example.bib

%\begin{thebibliography}{99}
%\end{thebibliography}

%%%%%%%%%%%%%%%%%%%%%%%%%%%%%%%%%%%%%%%%%%%%%%%%%%

%%%%%%%%%%%%%%%%% APPENDICES %%%%%%%%%%%%%%%%%%%%%

\appendix
\section{Numerical Schemes}
\label{ap:math}

\subsection{Steady state solution}
In this section we compute the SSS, $n_{\gamma 0}=n_{\gamma}\big[ \frac{\partial}{\partial t}=0 \big]$ of equation (\ref{eq:Kompaneets}). After the change of variables $n_{\gamma} = N n_c$ and $E = x kT_e$ equation (\ref{eq:Kompaneets}), in steady state, becomes: 
 
\begin{equation}
\frac{d^2 N}{dx^2} = - \frac{dN}{dx} + N \Big( \frac{2}{x^2} - \frac{2}{x} + \frac{c_2}{x^2} \Big) - \frac{1}{e^{x \frac{T_e}{T_s}} - 1} \ \ ,
\label{eq:SS}
\end{equation}

\noindent where $c_2 = \frac{m_e c^2}{kT_e N_{esc}(x)}$ and $n_c = \frac{2 \pi m_e t_c kT_{e0} 3a^2}{h^3 [(a+L)^3 - a^3] } $. Equation (\ref{eq:SS}), which is dimensionless after the change of variables, is solved as a boundary value problem, using a finite difference scheme. The derivatives in equation (\ref{eq:SS}) were replaced by second-order accurate central differences formulas. The discretised equation on an arbitrary mesh is:   

\begin{equation}
\begin{split}
N^{j-1} \underbrace{\Big( \frac{1}{\delta x^2}-\frac{1}{2\delta x} \Big)}_{L} + N^j \underbrace{\Big(  -\frac{2}{\delta x^2} -\frac{2}{x_j^2} + \frac{2}{x_j} - \frac{c_2}{x_j^2}  \Big)}_{D(x_j)} \\ + N^{j+1} \underbrace{\Big(  \frac{1}{\delta x^2} + \frac{1}{2\delta x_j} \Big)}_{U} = \underbrace{- \frac{1}{e^{x_j \frac{T_e}{T_s}} - 1} }_{C(x_j)},
\label{eq:dSS}
\end{split}
\end{equation}

\noindent where $j=1,2,...,M-1$, with $M$ the number of mesh points and $\delta x$ the mesh size, i.e. step size of the method. (Note that although we use $\delta$ for the step size, it does not represent any kind of perturbation. The condition applied at the boundaries is naturally $N^{0}=N^{M}=0$. The solution of equation (\ref{eq:SS}) comes from solving a linear system that is formed using equation (\ref{eq:dSS}). In particular, if we write equation (\ref{eq:dSS}) for every consecutive triad of points on the mesh, we can form an $(M-1)\times (M+1)$ linear system. In matrix form this equation will be \textbf{A}$\times$\textbf{N} = \textbf{C}. Taking into account the boundary conditions, the first and last columns of $\textbf{A}$ can be incorporated in $\textbf{C}$ so now the new coefficient matrix \textbf{T} of the system is square $(M-1)\times(M-1)$. Matrix \textbf{T} is tri-diagonal with L,D,U (appearing in \ref{eq:dSS}) being its sub-diagonal, diagonal and super-diagonal elements respectively. We solved this system using Gaussian elimination with partial pivoting using the routine \textbf{$dgtsv$}  from the LAPACK library \citep{lapack} as implemented in scipy's package \textbf{$linalg$} \citep{linalg}.

\subsection{Solution of the linearized equation}
\label{ap:pert}

Initially, we derive the formulas that describe the fractional amplitudes of the perturbations $\delta \dot{H}_{ext}$ , $\delta T_e$ and $\delta T_s$ and discuss how they are connected to $\delta n_{\gamma}$. Following the reasoning and formulation of \citet{Kumar2014}, and using the first law of thermodynamics, we can write:

\begin{equation}
\frac{3}{2} k \frac{\partial T_e}{\partial t} = \dot{H}_{ext} - <\Delta\dot{E}> \ \ ,
\label{eq:law}
\end{equation}

\noindent where $<\Delta\dot{E}>$ can be viewed as the work rate produced per particle in an ideal gas, here electrons in the medium, $\dot{H}_{ext}$ is the heat offered to the system per unit time per particle, and $\frac{3}{2} k (\partial T_e/\partial t)$ is the change in the internal energy per particle per unit time. The term $<\Delta\dot{E}>$ for the Comptonization process is identified as the Compton cooling rate per electron and is given by:

\begin{equation}
<\Delta\dot{E}> = \int\displaylimits_{E_{min}}^{E_{max}} (4kT_e - E)\frac{E}{m_e c} n_{\gamma} \sigma_T dE \ \ ,
\label{eq:cooling}
\end{equation}

\noindent where $E_{min}$ and $E_{max}$ are arbitrary limits. The bolometric luminosity of the NS surface, is  $4\pi a^2 \sigma T_s^4 $, where $\sigma$  is the Stefan Boltzman constant. According to \citet{Lee2001} and \citet{Kumar2014} it is valid to assume that a fraction $\eta$ of the output Comptonized photons return to the NS, increasing its temperature and thus its luminosity. Based on this effect we can model the luminosity of the NS as a superposition of a luminosity of a black body component with an inherent temperature $T_{s,inh}$,  and a fraction, $\eta$, of the output spectrum. In that sense we can write:

\begin{equation}
4\pi a^2 \sigma T_s^4 = 4\pi a^2 \sigma T^4_{s,inh} + \eta V_c \int \frac{n_{\gamma}}{t_c N_{esc}(E)} E dE \ \ ,
\label{eq:disc}
\end{equation}

\noindent where $V_c = (4/3)\pi[ (a+L)^3 - a^3]$ is the volume of the medium. Hereafter $\eta$ will be referred to as the feedback parameter; it is obvious from equation (\ref{eq:disc}) that there is an upper limit $\eta_{max}$ for $\eta$ that is obtained if we apply $T_{s,inh}=0$.  Assuming that the inherent NS temperature, $T_{s,inh}$, does not undergo oscillations then $\delta T_s$ is the result of the oscillating photon density, $\delta n_{\gamma}$, that affects the observed temperature, $T_s$, through feedback.  

 The amplitudes $\delta T_e$, $\delta T_s$ can be obtained by linearizing equations (\ref{eq:law}) and (\ref{eq:disc}) after substituting the perturbed terms. The derived formulas for the amplitudes, after the proper change of variables, are: 

\begin{equation}
\delta T_e = \frac{\dot{H}_{ext0} \dot{\delta H_{ext}} + \frac{n_c \sigma_{KN}(x) (kT_{e0})^3 }{m_ec}(\int x^2 N \delta n_{\gamma} dx-4\int x N \delta n_{\gamma} dx)    }{-\frac{3}{2}i\omega kT_{e0} + \frac{4(kT_{e0})^3 n_c \sigma_{KN}(x) }{m_ec }\int xNdx } \ \ ,
\label{eq:DTe}
\end{equation}

and 

\begin{equation}
\delta T_s = \frac{k^4 \eta V_c n_c (kT_{e0})^2}{16 \pi a^2 \sigma (kT_{s0})^4 t_c } \int \frac{x N \delta n_{\gamma}}{N_{esc}(x)} dx \ \ .
\label{eq:DTd}
\end{equation}

 In \ref{eq:DTe}, $\dot{H}_{ext0}$  is the, assumed constant, external heating rate when the corona is at thermal equilibrium. In order to calculate this term we must consider the equilibrium solution ($\frac{\partial T_e}{\partial t} = 0$), of equation (\ref{eq:law}), which leads to: 

\begin{equation}
\dot{H}_{ext0} = \frac{(kT_{e0})^3 n_c \sigma_{KN}(x) }{m_ec} \Bigg( 4\int xNdx - \int x^2Ndx \Bigg) \ \ .
\label{eq:Hext}
\end{equation}

 With equation (\ref{eq:Hext}), we complete the set of equations that are used to construct the model. Finally, we must note that the integrals that appear in all of the aforementioned formulas must be computed numerically and the integration limits are arbitrary in general.

Equation (\ref{eq:CODE}), which already is in dimensionless form, was discretized using second-order accurate central difference formulas, to become:

\begin{equation}
{\scriptsize
\begin{split}
\delta n_{\gamma}^{j-1} \underbrace{ \Big( -\frac{1}{\delta x^2} + \frac{1}{2 \delta x} + \frac{dN}{dx} \Bigg|_{x_j}\frac{1}{N \delta x}  \Big) }_{L(x_j)}  + \delta n_{\gamma} ^j \underbrace{\Big( \frac{2}{\delta x^2} + \frac{1}{N} \frac{1}{e^{x_j \frac{T_e}{T_s}}-1} - \frac{ic_5}{x_j^2}\Big)}_{D(x_j)}  \\ + \delta n_{\gamma}^{j+1} \underbrace{\Big( -\frac{1}{\delta x_j^2} - \frac{1}{2 \delta x} - \frac{dN}{dx}\Bigg|_{x_j}\frac{1}{N \delta x}  \Big)}_{U(x_j)}  = \\
 \underbrace{ \delta T_e \Big(  - \frac{2}{x_j^2} +\frac{1}{N}\frac{d^2N}{dx^2} \Bigg|_{x_j} \Big) + \delta T_s \frac{1}{N} \frac{T_{e0}}{T_{s0}} \frac{x_j}{\big( e^{x_j \frac{T_{e0}}{T_{s0}}}+e^{-x_j\frac{T_{e0}}{T_{s0}}}-2  \big)}}_{C(x_j)} \ \ ,
\label{eq:dCODE}
\end{split}
}
\end{equation}

Equation (\ref{eq:dCODE}) represents a linear system of equations, where the left hand side can be used to construct a tri-diagonal coefficient matrix with $L(x_j)$, $D(x_j)$ and $U(x_j)$ the sub diagonal, diagonal and super diagonal elements respectively. For simplicity we write equation (\ref{eq:dCODE}) as: 

\begin{equation}
\begin{split}
\delta n_{\gamma}^{j-1} L(x_j)   + \delta n_{\gamma} ^j D(x_j)  + \delta n_{\gamma}^{j+1} U(x_j) = C_1(x_j) \delta T_e + \\
  C_2(x_j)\delta T_s \  ,
\label{eq:dCODE_simple}
\end{split}
\end{equation}

\noindent where $C_1(x_j)$ and $C_2(x_j)$ can be found by comparing equation (\ref{eq:dCODE_simple}) to (\ref{eq:dCODE}). Assuming $M$ mesh points, just as in equation (\ref{eq:dSS}) so that $j=1,2,...,M$, the left hand side of equation (\ref{eq:dCODE_simple}) can be represented by a product of a tri-diagonal array of coefficients, with $L$,$D$ and $U$ being its sub diagonal, diagonal and super diagonal elements respectively, and a column array of the unknown values of $\delta n_{\gamma}$ on the grid.

On the right hand side of equation (\ref{eq:dCODE_simple}) the terms $\delta T_s$ and $\delta T_e$ are integral functions of $\delta n_{\gamma}(x)$ thus we can write for $\delta T_e$:

\begin{equation}
\begin{split}
 \delta T_e = k_0(x) + k_1(x) \int Q_1(x) \delta n_{\gamma} dx + k_2(x) \int Q_2(x) \delta n_{\gamma}dx \ ,
\label{eq:dTe_simple}
\end{split}
\end{equation}

\noindent where $k_1$,$k_2$,$k_3$, $Q_1$ and $Q_2$ can be found by comparing equation (\ref{eq:DTe}) to (\ref{eq:dTe_simple}). For $\delta T_s$ we can write:

\begin{equation}
\begin{split}
 \delta T_s = p_1 \int Q_3(x)  \delta n_{\gamma} dx   \ ,
\label{eq:dTs_simple}
\end{split}
\end{equation}

\noindent where $p_1$ and $Q_3$ can be found by comparing equation (\ref{eq:DTd}) to (\ref{eq:dTs_simple}). We can now use the composite 3/4 Simpson rule in order to replace the integrals in equations (\ref{eq:dTe_simple}) and (\ref{eq:dTs_simple}) with analytic expressions. The definite integral of a random function $F$ whose analytic form is unknown, using second order Lagrange polynomials, can be written as

\begin{equation}
\begin{split}
 \int_a^b F(x)dx = \frac{\delta x}{3} \sum_{j=1}^{M/2} [F(x_{2j-2}) + 4F(x_{2j-1}) + F(x_{2j})]  \ .
\label{eq:simpson}
\end{split}
\end{equation}

In our case, the integrals in equations (\ref{eq:dTe_simple}) and (\ref{eq:dTs_simple}), that we want to approximate, have the general form  $I_i = \int Q_i(x)\delta n_{\gamma}(x)dx$ where $i$ is an index that denotes a different function. According to equation (\ref{eq:simpson}) our integrals $I_i$ can be written as a weighted sum of the values of $\delta n_{\gamma}(x_j)$ on the grid as:

\begin{equation}
\begin{split}
 \ \ \ \ \ \ \ \ \ \ \ \ I_i = \frac{\delta x}{3} \Big[ Q_i(x_0)\delta n_{\gamma}(x_0) +  Q_i(x_{M})\delta n_{\gamma}(x_{M})\Big] + \\ 
\frac{4 \delta x}{3} \sum_{\substack{j=1\\\mod{j/2}\neq 0}}^{M-1} Q_i(x_j)\delta n_{\gamma}(x_j) \\ +  \frac{2\delta x}{3} \sum_{\substack{j=1\\\mod{j/2}= 0}}^{M-2} Q_i(x_j)\delta n_{\gamma}(x_j) 
 \ .
\label{eq:I_i}
\end{split}
\end{equation}

Using equations (\ref{eq:dTe_simple}), (\ref{eq:dTs_simple}) and (\ref{eq:I_i}) the amplitudes $\delta T_e$ and $\delta T_s$ can be substituted to the right hand side of equation (\ref{eq:dCODE_simple}), as the sum of the values of $\delta n_{\gamma}(x_j)$ on the entire grid. The final form of the coefficient array, corresponding to the solution of the system of equations formed by equation (\ref{eq:dCODE_simple}), will not be tri-diagonal anymore. The advantage of this mathematical technique is the fact that no initial assumptions are needed for $\delta T_e$ and $\delta T_s$, and also there is no need of an iterative scheme that requires the aforementioned amplitudes to converge in order to obtain a final solution.

% Don't change these lines
\bsp	% typesetting comment
\label{lastpage}
\end{document}